\documentclass[sigconf]{acmart}

\AtBeginDocument{%
  \providecommand\BibTeX{{%
    \normalfont B\kern-0.5em{\scshape i\kern-0.25em b}\kern-0.8em\TeX}}}

\newcommand{\QUOTE}[1]{\textsf{\textit{``#1''}}}
\newcommand{\proxy}[1]{\textsc{#1}}

\copyrightyear{2024}
\acmYear{2024}
\setcopyright{acmlicensed}\acmConference[DIS '24]{Designing Interactive
Systems Conference}{July 1--5, 2024}{IT University of Copenhagen, Denmark}
\acmBooktitle{Designing Interactive Systems Conference (DIS '24), July 1--5,
2024, IT University of Copenhagen, Denmark}
\acmDOI{10.1145/3643834.3661556}
\acmISBN{979-8-4007-0583-0/24/07}

\usepackage[english]{babel}
\usepackage{hhline}
\usepackage{multirow}
\usepackage{makecell}
\usepackage{graphicx}
\usepackage{tabularx}
\usepackage[ruled,vlined]{algorithm2e}
\usepackage{listings}
\usepackage{caption}
\usepackage{subcaption}

\usepackage{soul}
\makeatletter

\newcommand\stB[1]{\@bsphack\@esphack}
\renewcommand\st[1]{\@bsphack\@esphack}
\makeatother

\usepackage{soul}

\def\name{SoundShift}

\author{Ruei-Che Chang}
\email{rueiche@umich.edu}
\affiliation{
 \institution{University of Michigan}
 \city{Ann Arbor, MI}
 \country{USA}
}

\author{Chia-Sheng Hung}
\email{yoyung0809@gmail.com}
\affiliation{
 \institution{National Taiwan University}
 \city{Taipei}
 \country{Taiwan}
}

\author{Bing-Yu Chen}
\email{robin@ntu.edu.tw}
\affiliation{
 \institution{National Taiwan University}
 \city{Taipei}
 \country{Taiwan}
}

\author{Dhruv Jain}
\email{profdj@umich.edu}
\affiliation{
 \institution{University of Michigan}
 \city{Ann Arbor, MI}
 \country{USA}
}

\author{Anhong Guo}
\email{anhong@umich.edu}
\affiliation{
 \institution{University of Michigan}
 \city{Ann Arbor, MI}
 \country{USA}
}

\begin{document}

\title{SoundShift: Exploring Sound Manipulations for Accessible Mixed-Reality Awareness}

\renewcommand{\shortauthors}{Chang et al.}

\begin{abstract}
Mixed-reality (MR) soundscapes blend real-world sound with virtual audio from hearing devices, presenting intricate auditory information that is hard to discern and differentiate. 
This is particularly challenging for blind or visually impaired individuals, who rely on sounds and descriptions in their everyday lives.
To understand how complex audio information is consumed, we analyzed online forum posts within the blind community, identifying prevailing challenges, needs, and desired solutions.
We synthesized the results and propose \emph{{\name}} for increasing MR sound awareness, which includes six sound manipulations: \proxy{Transparency Shift}, \proxy{Envelope Shift}, \proxy{Position Shift}, \proxy{Style Shift}, \proxy{Time Shift}, and \proxy{Sound Append}. 
To evaluate the effectiveness of {\name}, we conducted a user study with 18 blind participants across three simulated MR scenarios, where participants identified specific sounds within intricate soundscapes.
We found that {\name} increased MR sound awareness and minimized cognitive load. 
Finally, we developed three real-world example applications to demonstrate the practicality of {\name}.
\end{abstract}




\settopmatter{printfolios=true}

\begin{teaserfigure}
  \vspace{-.5pc}
  \includegraphics[width=\linewidth]{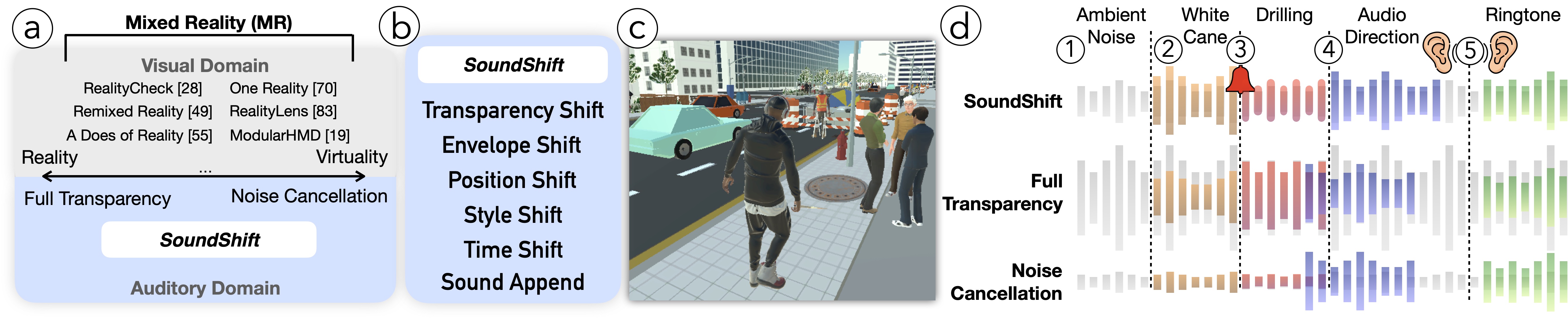}
  \vspace{-1.5pc}
  \caption{
  We present \textit{{\name}}, a concept to manipulate sounds to improve mixed-reality awareness. 
  (a) {\name} is situated in the auditory Reality-Virtuality Continuum with full transparency and noise cancellation as two ends, and comprises 
  (b) six sound manipulators, which are \proxy{Transparency Shift}, \proxy{Envelope Shift}, \proxy{Position Shift}, \proxy{Style Shift}, \proxy{Sound Append}, and \proxy{Time Shift}. 
  (c) In a scenario, a BVI user navigates a busy street with a white cane and audio directions. They sometimes may receive ringtones and pass by construction sites with drilling noises.
  (d.1) \proxy{Transparency Shift} makes the auditory transparency half to suppress nuanced noises while retaining real-world awareness.
  (d.2) \proxy{Envelope Shift} increases the white cane sounds to make them distinctive.
  (d.3) \proxy{Sound Append} plays an earcon to signal the danger, and \proxy{Style Shift} applies a low-pass filter to make drilling noise less sharp to hear.
  (d.4) \proxy{Time Shift} delays the audio directions when they conflict with drilling noises.
  (d.5) \proxy{Position Shift} places ringtone on the right and audio directions on the left to increase distinguishability. 
  }
  \Description{
    Figure 1 An overview of SoundShift system
    Figure 1a
    In the lower part of the figure, SoundShift is in the middle, and Full Transparency and Noise Cancellation are at the two ends on the left and right sides. In the upper part of the figure, there is text visual domain as the main topic, and virtuality is pointed to the right while reality is pointed to the left. Also, there are some related works below it, such as RealityCheck, One Reality, Remixed Reality, A Dose of Reality and ModularHMD. 
    Figure 1b
    SoundShift include six sound manipulators: Transparency Shift, Envelope Shift, Style Shift, Position Shift, Time Shift, and Sound Append. 
    Figure 1c
    A screenshot of Unity 3D world. A man wearing a black jacket is navigating the street with the white cane. On the street, there are cars whizzing by and crowds talking to each other. 
    Figure 1d
    This is the figure showing sound waveforms of three study conditions: SoundShift, Full Transparency, and Noise Cancellation. Details can be referred to the image captions.
  }
  \label{fig:teaser}
  \vspace{.5pc}
\end{teaserfigure}

\maketitle

\section{Introduction}\label{intro}
Mixed reality (MR) is becoming more pervasive nowadays, where real-world (RW) and virtual-reality (VR) visual elements blend and interact with each other in real time, offering users seamless access to both.
Visual descriptions \cite{mobilear,omniscribe} or acoustic cues \cite{vrbubble, acoustic_game, collabally} could make visual MR more accessible to people who are blind or visually impaired (BVI) who rely heavily on sounds in their everyday lives. 
However, there is little discussion on the accessibility of auditory MR, where sounds come from different sources, from RW content to virtual audio presented by hearing devices, which may conflict with each other.

The conflict of sounds could cause confusion and high cognitive load to BVI people and potentially lead to missing crucial information.
For instance, imagine walking down a busy street with ambient and crowd noises while receiving navigation instructions and participating in a virtual call. 
It might be difficult to shift one's focus across different audio applications.
Such situations will be increasingly common, as MR applications become more pervasive and have a growing integration of virtual sounds in our lives such as virtual meetings, text-to-speech applications, broadcast, and music/entertainment. 
Furthermore, the necessity of having descriptions or acoustic cues for non-visually accessing MR would also impose another audio layer on BVI people, and the higher quality and fidelity of synthesized voice or virtual sounds may be hard to discern from RW counterparts.
In this work, we aim to investigate the following question: \emph{How to manipulate sounds to enhance sound awareness in a complex MR audio environment for BVI people?}

To understand the current practices in consuming complex audio information, we first conducted a content analysis from active online forums within the BVI community, where we collected posts and comments about the prevailing scenarios, challenges, and potential solutions for consuming complex audio information. 
We found that many scenarios require the consumption of RW\footnote{RW sounds refer to sounds from RW environments, such as crowd, speaker, television.} and VR\footnote{VR sounds refer to sounds rendered through users' hearing devices.} sounds, such as navigating a busy street with audio directions, playing an instrument following music tutorials, and consuming screen reader feedback and other audio applications. 
On the other hand, BVI people expressed their desires to manipulate sounds and proposed ad-hoc solutions, such as distributing sound sources to different devices to allow the consumption of multiple sounds in parallel, adjusting sound characteristics to make sounds distinctive, or customizing existing sound libraries of applications.
We synthesized our findings and proposed \emph{\text{\name}} to enhance users' perception and awareness of sounds in mixed reality environments. This approach incorporates six sound manipulation techniques: \proxy{Transparency Shift}, \proxy{Envelope Shift}, \proxy{Position Shift}, \proxy{Style Shift}, \proxy{Time Shift}, and \proxy{Sound Append}.

To understand how {\name} manipulations can affect the perception of MR sounds, we conducted a user study with eighteen BVI participants who experienced the three simulated scenarios and identified sounds, including a \emph{RW-Focused} scenario when navigating a busy street, a \emph{VR-Focused} scenario when focusing on audiobook, and a \emph{Fully-Mixed} scenario when attending a hybrid conference.
In each scenario, we applied pre-defined sound manipulations on real-world and virtual sounds to enhance the perception of the mixed-reality soundscape.
We compared {\name} with the other two ends of the auditory Reality-Virtuality continuum \cite{milgram1994taxonomy,larsson2010auditory,mcgill}: full acoustic transparency that enhances the presence of the real world, and noise cancellation that enhances the immersion of the virtual reality (Figure \ref{fig:teaser}a). 

Our results showed that sound manipulations significantly improved BVI people's ability to perceive and manage sound information compared to full transparency and noise cancellation in our three simulated scenarios.
The conditions also had varying effects on participants' performance across the three scenarios.
{\name} also significantly reduced participants' cognitive load in perceiving and managing sounds compared to full transparency and noise cancellation.
Additionally, participants shared ideas for how they would further customize the sound manipulations in each scenario. 
Our evaluations demonstrate that {\name} effectively improved MR sound awareness for BVI people.

To demonstrate the generalizability and practicality of {\name}, we further developed three real-world example applications based on our content analysis on online posts and user customization mentioned in our study:
\textit{(i)} an audio-adaptive online meeting web application that addresses the conflict between screen reader sounds and meeting conversations, 
\textit{(ii)} a mixed-reality content-aware image exploration application that provides stylized and spatialized audio feedback based on real-world and virtual content, and
\textit{(iii)} a mobile navigation application that analyzes real-world and virtual sound events to identify opportune moments for delivering audio directions.

In summary, our work contributes: 
\begin{itemize}
    \item The concept of {\name} to make MR sound awareness accessible for BVI people, through six sound manipulators derived from our content analysis on BVI forums.
    
    \item An instantiation of the six sound manipulators and three simulated scenarios across the Reality-Virtuality continuum in Unity.
    
    \item Results from a user study with eighteen BVI people to provide empirical evidence of significantly enhanced sound awareness through {\name} manipulations.
    
    \item Three real-world example applications to demonstrate the generalizability and practicality of {\name}.

\end{itemize}

\section{Related work}
Our work was motivated and situated from prior work on mixed reality, MR accessibility, and soundscape personalization.

\subsection{Blending Real and Virtual World} \label{RW_MRblend}
Several prior works have explored blending Reality and Virtuality \cite{milgram1994taxonomy} (Figure \ref{fig:teaser}a) to achieve novel interactions by leveraging otherwise unavailable benefits from the other. 

In Augmented Virtuality (AV), which blends RW elements into virtual experiences, A Dose of Reality \cite{doseofreality} highlighted the benefits of incorporating RW components, like using a visible physical keyboard in VR to improve typing. 
RealityLens \cite{realitylens} introduced techniques for users to personalize integrating RW visual regions into VR. 
ModularHMD \cite{modularhmd} promoted RW awareness via peripheral views through HMD configurations. 
RealityCheck \cite{realitycheck} delved into visual blending methods for fusing physical and virtual details. 
Further research has also pinpointed key aspects for fluidly embedding RW digital data (e.g., smartphone alerts) into VR, examining optimal intervention times \cite{opportune}, notification display methods \cite{notifivr, immersivenoti}, and proper positions for notifications \cite{notiposition}.

As for Augmented Reality (AR) \cite{milgram1994taxonomy}, One Reality \cite{onereality} provided a conceptual framework that embodied the incremental levels of mixed-reality interactions from physical to virtual worlds. 
Remixed Reality \cite{remixedreality} characterized the manipulations of time and space that allow users to experience live 3D reconstruction as a novel form of MR.
Given the growing works exploring blended interactions, VRception \cite{vrception} was thus developed as a toolkit to facilitate the rapid prototyping of MR interactions. 

Beyond visual blending, tangible real-world objects serve as tactile proxies for objects in VR, which create enhanced haptic experiences  \cite{everydayobject, annexingreality, subsreality}. 
While numerous prior works have investigated blending RW and VR in visual and haptic domains, (un)blending sounds in MR is still under-explored, which is essential to creating accessible awareness for BVI people.

\subsection{Soundscape Formation and Personalization} \label{soundscape_personalization}
Augmented Audio Reality (AAR) integrates virtual sounds into RW soundscapes and is prevalent in our everyday lives \cite{sixtypes}. 
Larsson et al. \cite{larsson2010auditory} posited that AR or AV in the concept of the Reality-Virtuality continuum can also adapt to auditory domains.
McGill et al. \cite{mcgill} found that acoustically transparent headphones increase the sense of presence of reality while noise-canceling headphones diminish it.
Advancements in noise-canceling technologies (also known as soundscaping technologies \cite{hagood2011quiet}), have transformed the headphone user experience and emerged as a significant research field \cite{soundscapesurvey}.
However, Haas et al. \cite{curation} described the limitations of soundscaping technologies: \emph{``current personal audio technology is not designed in a way that it allows users to handle their social context satisfactorily. Furthermore, information acquisition is made more difficult and users often have to make a choice between the surrounding acoustic environment and their own content and media.''} 
Though prior research \cite{cliffe2021materialising, mamuji2005attentive, sawhney2000nomadic, bederson1995audio} or the current commercial headphones support adapting the audio based on users' environment, users still expressed desires to steer and personalize their soundscape in a fine-grained manner \cite{curation} (e.g., blocking certain unwanted sounds, masking environmental sounds).
This is particularly crucial for BVI individuals, who may have different vision levels and reliance on auditory cues, resulting in varying needs for curating soundscapes in different contexts \cite{rychtarikova2015blind}.
Recently, advancements in AI within the auditory domain have enhanced AAR and opened avenues for tailoring soundscaping technologies to users' needs. 
For example, several works have facilitated intelligent sound extraction \cite{semantichearing,veluri2023real,clearbuds}, enabling users to selectively filter or emphasize specific sounds.
Furthermore, sounds can be intelligently adapted to the user's activities, offering seamless experiences. This includes altering music based on driving contexts to improve in-car music experiences \cite{soundsride}, or integrating audio notifications with ongoing music to reduce disruption and annoyance \cite{maringba}.

Inspired by these needs and trends in personalizing soundscapes, we explore more fine-grained sound manipulations tailored specifically for BVI individuals, who, as experts in audio technologies, may have a unique reliance and strategies on manipulating sounds.
We seek to understand how these fine-grained manipulations could enhance sound awareness for BVI people in various MR scenarios, and how BVI people would perceive and customize these sound manipulations.

\subsection{Accessibility of Mixed Reality}
In recent years, the World Wide Web Consortium (W3C) has established guidelines for MR accessibility \cite{xraccessibiltiy} to encourage MR developers to consider the diverse needs of people with different abilities as a primary concern rather than an ``afterthought'' once the technology has matured \cite{a11ybydesign}.
Efforts to enhance MR for people with disabilities include visual \cite{soundvizvr} or haptic \cite{soundvra11y} alternatives for people who are deaf or hard of hearing (DHH), and simplified interaction techniques for people with motor impairments \cite{limitedmobility, nearmi}.

BVI people, on the other hand, often face challenges in fully experiencing MR.
Tools like SeeingVR \cite{seeingvr} have been developed to enhance visual awareness for people with visual impairments and ensure accessibility to VR content, and Herskovitz et al. \cite{mobilear} concluded a design space for AR tasks and made them accessible via corresponding verbal feedback. 
Furthermore, various hardware devices have been developed to enhance the haptic experiences, such as cane simulations in VR for accessible navigation and a better sense of immersion \cite{navstick, canetroller, d2007system, virtualcane, lecuyer2003homere}.
As for audio domains in MR scenarios such as navigation, acoustic maps \cite{acousticmap,nair2022uncovering,acousticmapCSCW23} or spatial audio \cite{acousticmap,nair2022uncovering,acousticmapCSCW23, gpssound,soundscape,blindsquare} convey the area information to BVI individuals in an acoustic form that helps them construct mental maps in VR.
Also, VRBubble \cite{vrbubble} employed sound representations to bolster BVI peripheral awareness during social interactions, 
and OmniScribe \cite{omniscribe} made the immersion of 360° videos accessible to BVI people by rendering traditional audio descriptions spatially based on the orientation of BVI users.

From the trend that prior works proposed different auditory solutions to address certain accessibility challenges, it is expected that an accessible mixed reality for BVI people would entail much more complex audio information. 
Our work, therefore, aims to help increase BVI people's awareness of complex MR sounds by exploring effective sound manipulations.

\section{Understanding practices of consuming complex sounds}
In this work, we aim to investigate: \emph{How to manipulate sounds to enhance sound awareness in a complex MR audio environment for BVI people?}
To answer this question, we first need to understand the current challenges, needs, and practices that BVI people consume complex audio information. 

\subsection{Method}
We conducted a content analysis from active online forums within the BVI community. 
First, two researchers communicated synchronously over Zoom and reviewed posts and their comments in online forums, including AppleVis \cite{applevis}, and the Blind and Visually Impaired Community on Reddit \cite{blindreddit}. 
We started by filtering the posts by keywords such as ``sound'', ``audio'', ``headphone'', ``challenge'', and ``scenario.'' 
We reviewed posts from 2023 backward until we had collected over 1000 posts, ultimately reaching back to 2021.
Then, we eliminated off-topic posts, such as social activity recruitment, debugging devices, and casual conversations. 
Ultimately, we collected 100 posts and comments highly associated with the difficulties or needs of sound consumption. 
Then, we used thematic analysis \cite{clarke2015thematic} to analyze the data. 
Together, we reviewed and coded the posts and comments in an online spreadsheet and had discussions to reach an agreement on the major themes described below.

\subsection{Findings}
We organized our findings by reporting the scenarios of consuming complex audio information, the challenges, and potential solutions mentioned in the posts. 
In the below sections, RW sounds are those sourced from real-world objects, whereas VR sounds are auditory outputs generated through hearing devices.

\subsubsection{Everyday Scenarios to Consume Complex Audio Information}
\label{need_for_RW_VR_sounds}
RW or VR environment can create a complex ever-changing soundscape for BVI people. For instance, one stated their daily walk journey: \QUOTE{My normal daily 6 mile walk typically takes me through a range of environments - from busy roads with high levels of traffic noise, along quieter residential roads, [...]
to playing parks.}
Examples of complex VR soundscapes include when the screen reader overlaps with screen navigation feedback: \QUOTE{Is there a way to turn off the navigation sounds when using voiceover? I can customize these on my iPhone, so when I swipe I don't always get that clunking sound when encountering each item as I swipe}, or
when using a desktop interface while engaging in a virtual meeting: \QUOTE{When I am in a meeting, with a braille display and my watch, I do NOT want to keep hearing the click-click sounds VoiceOver makes when flicking right or left.} 
Several scenarios also required simultaneous consumption of RW and VR sounds, such as audio directions while navigating the street, screen reader feedback in a noisy environment, or playing instruments with music on the hearing device.

\subsubsection{Retaining Real-World Awareness in Noisy Environments}\label{retain_RW_awareness}
As mentioned, consuming both RW and VR sounds is common and inevitably creates several challenges for BVI people. 
RW noisy environments can overwhelm BVI people due to the difficulty in discerning important sounds out of noises, as one consumed Siri's sound effects: \QUOTE{[...] they are difficult to hear in noisy environments even if you don't have a hearing problem.}
Although noise-cancellation headphones can block out noises, BVI people still want to retain awareness of their surroundings:
\QUOTE{I want to be able to hear clearly what is around me at the same time I’m listening through them.}
From the online forums, bone-conduction headphones are frequently noted as a potential alternative by BVI people; however, there is a risk that VR sounds may be overshadowed by their RW counterparts.
Instead, one suggested dynamically adjusting virtual sound volume in accordance with the prevailing levels of RW sounds:
\QUOTE{You would always be aware of ambient sound while you were wearing them, and I liked the idea that volume could be set to adjust automatically depending on the noise around you.}
This idea echoes prior research on developing audio-adaptive systems for contextualized interactions \cite{cliffe2021materialising, mamuji2005attentive, zimmermann2008listen}.

\subsubsection{Adjusting Sound Characteristics of Important Sounds in Conflict with Each Other}
\label{adjust_sound_characteristics}
Aside from conflicting with ambient noises, important sounds may conflict with each other and cause distraction or interference.
For instance, one comment said: 
\QUOTE{When we are listening to music or watching shows we keep having what is going to play up-next, which sometimes cuts off the start of a track in the case of music. [...]
I know it flashes up for sighted users but I feel for voiceover users it is an irritant.}
Several posts also inquired about the possibility of selective turning the screen reader on/off for specific applications, or dynamically adjusting sound characteristics to make the screen reader distinctive.
\QUOTE{They are difficult to hear in noisy environments [...] The problem here is that the sounds don't cover a wide enough range in the audio spectrum. There should be a variety of frequencies from low frequency to high frequency to make sure they can be heard in any environment.}

\subsubsection{Distributing Sound Sources for Better Distinction}\label{distribute_sound_source}
Besides adjusting sound characteristics, we also observed BVI people's strategy to distribute sound sources to different places for better perception of multiple audio streams. 
For example, one suggested routing audio streams to different devices: \QUOTE{Ability to route music to my smart speakers while keeping VoiceOver on my phone. It's a deal breaker when having a party and you get to hear VoiceOver on your surround sound when grooving to some music.}
Moreover, distributing sounds in different ears could be another solution:
\QUOTE{Frequently, I will just use one of the Beats Flex earpieces so that I limit the sound from my phone to one ear. For example, if traffic was on my right side, I might only use the left earpiece to hear directions from my phone.}
These findings reveal the promise of spatial audio for conveying information from many sources in MR scenarios beyond conveying directional information in previous works \cite{acousticmap,nair2022uncovering,acousticmapCSCW23,collabally,omniscribe} or commercial apps \cite{soundscape,blindsquare}.

\subsubsection{Customizing or Augmenting Existing Sound Library}\label{online_forum_augment_sound}
To avoid undesired audio presentations, we also found users' desires to customize the existing sound library, as one said: 
\QUOTE{It would just be nice to have maybe a different ring or two instead of either the old phone, the car horn which I find obnoxious}
or configuring the screen reader: \QUOTE{It should be possible to configure VoiceOver to play a special sound to indicate a control type such as a button or play two distinct sounds to quickly indicate if a checkbox is checked or unchecked.}
Also, to avoid overwhelming textual information of the screen reader, BVI people desired a proper amount of audio information, where earcons could play a vital role: \QUOTE{When any notification comes in, VoiceOver announces that there is a notification, then reads the time, and then eventually goes silent ... This seems like the wrong behavior, giving redundant information and is pretty annoying.
}
Besides, BVI people imagined customizing the voice of screen readers to their close ones: \QUOTE{I do love the idea of someone being able to save their voice ... Leads me to wonder what might happen to that voice when the person passes. Would we want to hear our own words spoken in a departed loved one's voice?}
These findings reveal users' needs of customizing sounds for different granularity of audio information.

\subsection{Summary}
Our findings revealed different scenarios where BVI people encountered complex audio information. These scenarios had different focuses spanning across the real world (e.g., traffic), virtual sounds (e.g., screen reader), or a combination of both (e.g., listening to music from the hearing device and playing instruments). Furthermore, BVI people proposed their desires and solutions to better consume the complex audio information, including manipulating ambient noises while retaining real-world awareness, adjusting sound characteristics to make important sounds distinctive, distributing sound sources in different devices or locations, and having customizations on earcons or voice feedback in devices. These findings inspired the three mixed-reality scenarios in our study and {\name} manipulations described in the next section.

\section{{\name} Manipulations for Accessible Mixed Reality Awareness}\label{soundblender}
{\name} is a concept to enhance mixed-reality sound awareness, which includes six sound manipulations: \proxy{Transparency Shift}, \proxy{Envelope Shift}, \proxy{Sound Append}, \proxy{Time Shift}, \proxy{Position Shift}, and \proxy{Style Shift}. 
We hypothesize that {\name} manipulations can enhance the awareness of sounds for BVI people in mixed-reality environments.
In this section, we describe how content analysis results and prior works inspired each sound manipulator. 

\textit{\textbf{\textsc{Transparency Shift}} modulates the ambient sounds or noises to shift presence between RW and VR by varying acoustic transparency.}
Acoustic transparency is a common description for headphones that blend virtual audio with real-world sounds \cite{mcgill}.
Real-world noises can enhance the sense of real-world grounding and presence, as so-called ``primitive hearing'' \cite{ramsdell1978psychology}. 
An increased sense of presence was also found when a user wears acoustically transparent headphones than the noise-cancellation ones \cite{mcgill}.
This creates opportunities to balance the presence and awareness between RW and VR through the degree of acoustic transparency \cite{gilkey1995sense,murray2000presence, larsson2010auditory}. 
The concept is similar to Apple Adaptive Audio \cite{appleaudio} by interpolating different auditory transparency and noise cancellation (Figure \ref{fig:teaser}d.1).

\textit{\textbf{\textsc{Envelope Shift}} modifies the dynamic aspects of sounds, such as volume or pitch, affecting their perceived loudness and tonal characteristics over time.}
Fundamental characteristics of sounds, such as pitch, volume, and duration, impact sensation and perception in daily listening \cite{gaver1988everyday, ballou1987handbook}. 
Modifying these characteristics, as discussed in section \ref{adjust_sound_characteristics}, helps make sounds distinguishable.
As shown in Figure \ref{fig:teaser}d.2, \textsc{Envelope Shift} increases the volume of specific sound sources over ambient noises. 

\textit{\textbf{\textsc{Position Shift}} controls the locations of sound sources to enhance the sense of immersion.}
Spatial audio has been a long-standing research field in VR and an essential component to embody the sense of presence and immersion \cite{lombard1997heart,brinkman2015effect,andre2012sound}.
In sections \ref{distribute_sound_source}, BVI people also distributed sound sources for selective attention, facilitating awareness through different directions.
As shown in Figure \ref{fig:teaser}d.5, \textsc{Position Shift} places sound sources at different spatial locations.

\textit{\textbf{\textsc{Style Shift}} transforms the timbre and fidelity of sounds through various filters to modify their aesthetic and emotional impact.}
Section \ref{online_forum_augment_sound} highlights BVI individuals' suggestions for screen reader voice customization beyond the standard synthesis. This includes diverse sound filters (e.g., high, low pass, human, robotic, anime) to cater to varied preferences. As shown in Figure \ref{fig:teaser}d.3, \textsc{Style Shift} uses a low pass filter to soften sharp drilling noises.

\textit{\textbf{\textsc{Time Shift}} adjusts the timing of sounds to prioritize or deprioritize them within a soundscape, controlling auditory focus.}
Overlapping sounds are common (sections \ref{need_for_RW_VR_sounds}, \ref{retain_RW_awareness}, \ref{adjust_sound_characteristics}). 
\textsc{Time Shift} controls sound timing to prevent overlap, such as pausing or delaying virtual audio in MR when it conflicts with RW sounds. 
As shown in Figure \ref{fig:teaser}d.4, \textsc{Time Shift} delays audio directions during drilling noises.

\textit{\textbf{\textsc{Sound Append}} appends earcons for corresponding sound events.}
Earcons \cite{blattner1989earcons} signal object updates and deliver key audio information with minimal user attention.
In complex MR environments, earcons can signal RW and VR events to reduce cognitive load instead of describing everything using speech. As shown in Figure \ref{fig:teaser}d.3, an earcon indicates potential danger.

\section{user Study}\label{userstudy}
Our goals of the user study are to 
\textit{(i)} explore the effect of the proposed sound manipulations, and
to \textit{(ii)} explore user preferences and customizations on sounds in different MR scenarios.
We created simulated environments in Unity to control the playback of sounds and their characteristics for simulating the proposed sound manipulations. 
This method aims to achieve high-fidelity simulations to immerse BVI people in the simulated environments, foster their engagement in the tasks, and help them imagine the future of sound technologies for providing feedback. 
This also enabled us to iterate and refine our concepts and user requirements based on their experiences and feedback before committing resources to develop fully functional systems. 
Our method was inspired by how Wizard-of-Oz methods were used in prototyping novel interactions \cite{lee2009snackbot,suede}, and how user enactment was used to elicit feedback by providing future usage scenarios \cite{userenactment1,userenactment2,userenactment3}.

Prior research indicated that full acoustic transparency (FT) increases RW presence and awareness, while noise cancellation (NC) enhances VR presence and awareness \cite{mcgill} (Figure \ref{fig:teaser}a).  
We, therefore, posit them as the two ends of auditory Reality and Virtuality Continuum \cite{larsson2010auditory} for optimally augmenting sound awareness in their respective realms. 
Consequently, our study compared {\name} manipulations with these two established conditions, hypothesizing that participants can achieve the best performance with SS in mixed-reality settings by combining the best of both worlds.
Specifically, we aim to understand the following research questions in this study:

\begin{itemize}
\item[\textit{RQ1:}] How do sound manipulations affect participants’ performance compared to full transparency and noise cancellation?
\item[\textit{RQ2:}] How do the different scenarios with varying emphases on reality and virtuality affect participants’ performance?
\item[\textit{RQ3:}] How do the conditions affect participants’ performance differently across the scenarios?
\item[\textit{RQ4:}] How do sound manipulations affect participants' cognitive load compared to full transparency and noise cancellation?
\item[\textit{RQ5:}] How do participants describe their experiences and ways to further customize their soundscape for each scenario?
\end{itemize}

\subsection{Participants}
Through word-of-mouth and public recruitment posts, we recruited 18 BVI participants (10 M and 8 F), aged 20 to 41 (mean=29.0), with a deep experience of using sounds in their lives.
All had professional Orientation and Mobility (O\&M) training and various sound-related experiences in other domains, like playing instruments, being voice actors, participating in orchestras, and broadcasting. 
Twelve were blind since birth, while six lost their vision later in life (\autoref{tab:demographic}). 
We refer to our BVI participants as B1-B18 in the following sections.

\begin{table*}[h]
  \caption{Participant demographics information. O\&M refers to Orientation and Mobility.}
  \label{tab:demographic}
  \vspace{-1pc}
  \begin{center}
  \begin{tabular}{|l|l|l|l|l|}
    \hline
    \textbf{ID} & \textbf{Age} & \textbf{Gender} & \textbf{Self-Reported Visual Ability} & \textbf{Hearing Experience} \\
    \hline
    B1 & 20 & Female & Blind, since birth. Light perception.  & O\&M, Orchestra, Instrument  \\
    \hline
    B2 & 31 & Female & Blind, since birth. Light perception. & O\&M, Instrument \\
    \hline
    B3 & 41 & Female & Blind, since birth. Light perception. & O\&M, Instrument, Voice Actor, Broadcast \\
    \hline
    B4 & 22 & Female & Blind, later in life. Light perception.  & O\&M, Instrument \\
    \hline
    B5 & 33 & Male & Blind, since birth & O\&M, Broadcast\\
    \hline
    B6 & 24 & Female & Blind, since birth & O\&M, Instrument\\
    \hline
    B7 & 22 & Male & Blind, since birth. Light perception. & O\&M, Instrument\\
    \hline
    B8 & 32 & Male & Blind, later in life & O\&M, Instrument\\
    \hline
    B9 & 20 & Male & Blind, since birth & O\&M, Instrument\\
    \hline
    B10 & 24 & Male & Blind, later in life & O\&M, Orchestra, Instrument  \\
    \hline
    B11 & 33 & Female & Blind, later in life & O\&M, Instrument \\
    \hline
    B12 & 23 & Male & Blind, since birth & O\&M, Instrument \\
    \hline
    B13 & 35 & Male & Blind, since birth & O\&M, Instrument \\
    \hline
    B14 & 31 & Male & Blind, since birth & O\&M, Instrument\\
    \hline
    B15 & 38 & Male & Blind, since birth & O\&M\\
    \hline
    B16 & 29 & Female & Blind, since birth. Light perception. & O\&M, Instrument, Making audio descriptions\\
    \hline
    B17 & 21 & Female & Blind, later in life. Light perception. & O\&M, Instrument\\
    \hline
    B18 & 33 & Male & Blind, later in life. Light perception. & O\&M\\
    \hline

  \end{tabular}
  \end{center}
  \Description{}
  \vspace{-1pc}
\end{table*}

\subsection{Simulated Scenarios and Sound Manipulations}\label{walkthrough}
To fully explore the space of MR, we present three scenarios with different focuses on sound awareness: \textit{RW-Focused}, \textit{VR-Focused}, and \textit{Fully-Mixed}.
We simulated the three scenarios in Unity and used its built-in audio functions to prototype the six manipulators. 
In each scenario, there were four types of sounds, and participants were asked to complete sound identification tasks by pressing down specific keys on the keyboard upon hearing corresponding sounds.
Participants were instructed to engage in the scenarios by prioritizing certain types of sounds relevant to the scenario's objectives, such as concentrating on RW sounds in the \textit{RW-Focused} scenario, or both RW and VR sounds in the \textit{Fully-Mixed} scenario.

\begin{figure}[b!]
\begin{center}
\includegraphics[width=\linewidth]{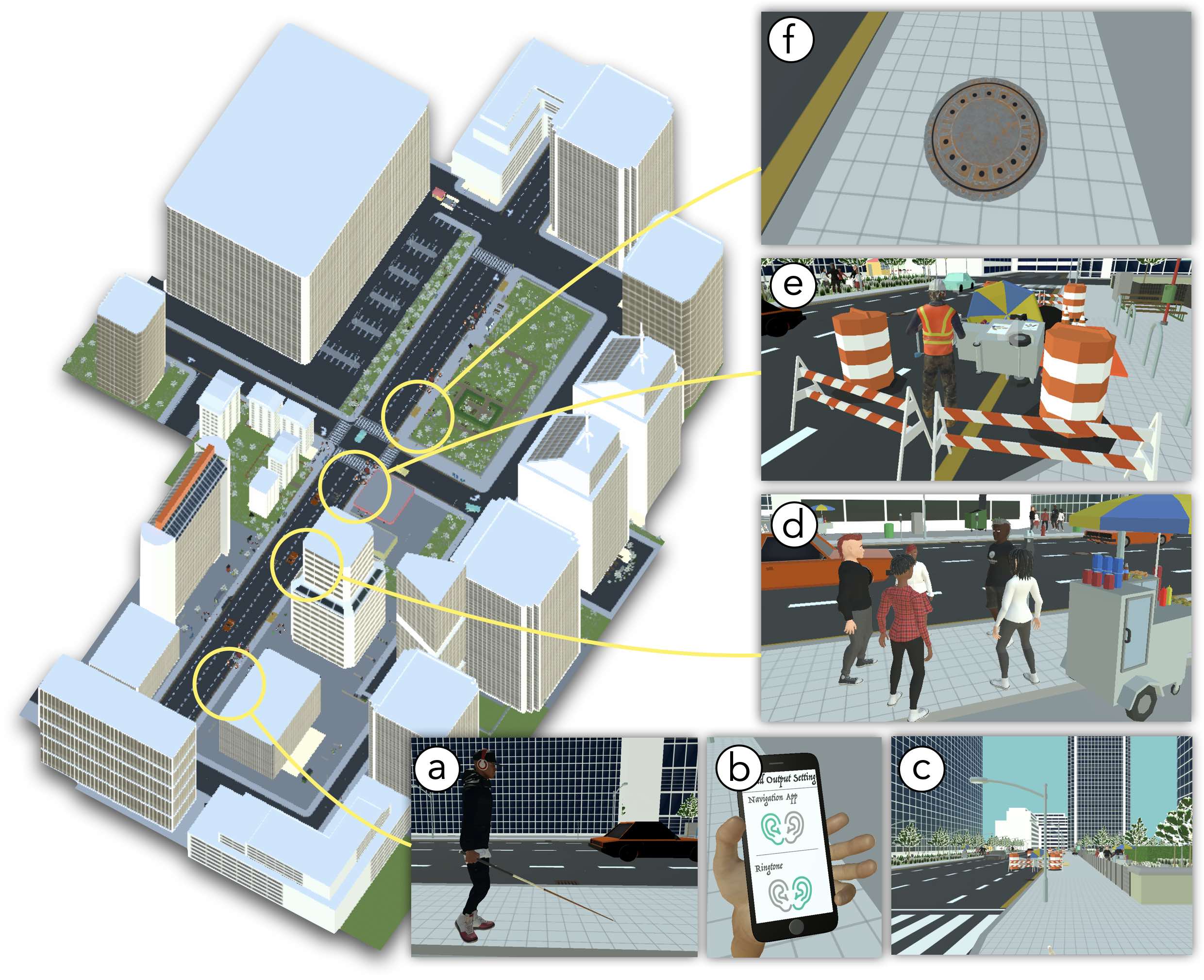}
\vspace{-1.8pc}
\caption{Simulated \textit{RW-Focused} Scenario. 
(a) The user's avatar in Unity wears a headphone and holds a white cane. 
(b) The user sets the sound output by placing audio directions on the left, ringtone on the right, and (c) navigates on the street.
(d) Several crowds, vendors, and passing cars along the street generate ambient noises, as well as (e) construction sites with drilling noises.
(f) While walking, the user might come across random manholes, causing the white cane's sound to change upon contact.
}
\label{fig:scene1}
\Description{Figure 2a
A man wearing a black jacket is navigating the street with the white cane. 
Figure 2b
The phone held by the man’s hands showing the sounds were spatialized on the mobile app.
Figure 2c
The first-person view of the man, including the road, street light, buildings and sidewalks.
Figure 2d
There are cars and crowds on the street and vendors on the sidewalk.
Figure 2e 
There are construction sites between the road and the sidewalk.
Figure 2f
A close look at the manhole.
}
\vspace{-1pc}
\end{center}
\end{figure}

\subsubsection{RW-Focused Scenario: Navigating on the street with a white cane and voice navigation guidance.}
\leavevmode 
\par
\textbf{Purpose:} We aimed to explore, in the RW-focused setting, whether sound manipulations can harmonize both RW and VR sounds to maintain user awareness, while the user is primarily focusing on real-world sounds.

\textbf{Motivation: } From our content analysis, navigating the road is an everyday routine for BVI people (Section \ref{need_for_RW_VR_sounds}), while they sometimes receive audio directions from navigation apps (Section \ref{need_for_RW_VR_sounds} and \ref{distribute_sound_source}). 
Thus, we simulated this scenario where the user focuses on the road conditions with occasional virtual audio presented, as detailed below.

\textbf{Scenario:}
This scenario simulates a user navigating with a white cane on a busy street full of crowd noises while using the smartphone navigation app to receive instructions (Figure \ref{fig:scene1}), such as ``turn left at the next intersection'', and receiving occasional phone ringtones during navigation.
In this scenario, the user periodically taps a white cane on the ground to detect surface changes, like manholes, which alter the cane's sound. They must also be alert to drilling noises from nearby construction sites, focusing mainly on real-world sounds for safety.

\textbf{Sound events and user required actions:}
For ambient noises, there are crowd noises randomly placed along the street, with occasional car noises to the user's left.
Participants were asked to press keys 1, 2, 3, and 4 to identify the white cane on a manhole, drilling, navigation, and ringtone, respectively.
The cane tapping sound occurs every 0.5 seconds, changing upon manhole contact for the next four taps. Each sound type has five random instances in our simulation, totaling 20 sound events to be identified.

\textbf{Sound manipulations:}
\begin{itemize}
\vspace{-.3pc}
\item \proxy{Transparency Shift}: applies half acoustic transparency (Figure \ref{fig:teaser}d.1).
\item \proxy{Envelope Shift}: prioritizes the four sound types by volume: white cane on the manhole, drilling, navigation instructions, and ringtone (Figure \ref{fig:teaser}d.2).
\item \proxy{Position Shift}: places navigation instructions on the left and ringtones on the right (Figure \ref{fig:teaser}d.5).
\item \proxy{Style Shift}: applies low pass filters to drilling noises to be audibly comfortable (Figure \ref{fig:teaser}d.3).
\item \proxy{Time Shift}: delays virtual sounds until after the RW sounds end (Figure \ref{fig:teaser}d.4).
\item \proxy{Sound Append}: appends a short earcon when getting close to a construction site (Figure \ref{fig:teaser}d.3).
\end{itemize}

\begin{figure}[b!]
\begin{center}
\includegraphics[width=\linewidth]{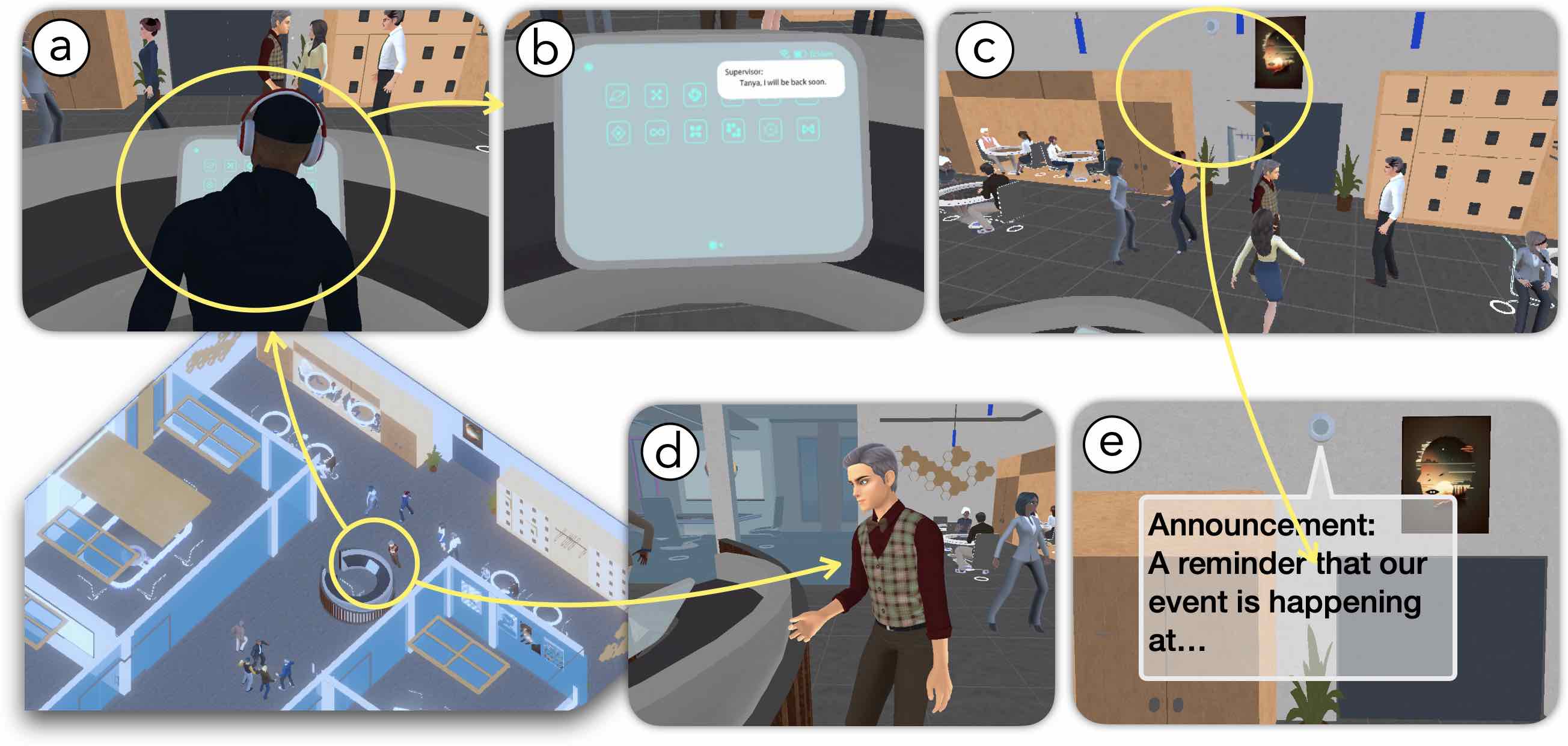}
\vspace{-1.8pc}
\caption{Simulated \textit{VR-Focused} Scenario. 
(a) The user’s avatar in Unity wears headphones, sits and works at the help desk, listens to an audio handbook, and (b) occasionally receives voice notes from a supervisor.
(c) In the environment, there are background noises when people walk around, talk to each other, and open/close the sliding door.
(d) People sometimes knock on the desk to get the user's attention.
(e) The speaker plays occasional public announcements on the front wall.
}
\label{fig:scene2}
\vspace{-1pc}
\Description{Figure 3
A global view of the Unity scene of VR-Focused scenario. 
Figure 3a
A third-person view looking at the back of the person, who wearing headphones sitting in front of the monitor.
Figure 3b
A monitor showing some app icons and a notification popup window is on the top right corner of the screen.
Figure 3c
The space in front of the help desk. There are crowds of people and there is a speaker on the wall and cabinets and round tables in the sides or the corners of the room.
Figure 3d
A person come to the help desk and knock on the edge of the table
Figure 3e
There is text showing “Announcement: A reminder that our event is happening at…”. This comes from the speaker on the wall.
}
\end{center}
\end{figure}

\subsubsection{VR-Focused Scenario: Consuming an audio handbook while working at the help desk.}
\leavevmode 
\par
\textbf{Purpose:}
We aimed to explore, in the VR-focused setting, whether sound manipulations can harmonize both RW and VR sounds to maintain user awareness, while the user primarily focuses on the virtual audio.

\textbf{Motivation:} From our content analysis, there are examples of focusing on virtual audio in a noisy environment (e.g., screen readers in section \ref{online_forum_augment_sound}).
Thus, we simulated this scenario where the user focuses on virtual tasks with occasional RW sounds presented, as detailed below.

\textbf{Scenario:}
This scenario depicts a user at a help desk, learning from an audio handbook as a new employee and responding to occasional knocks for attention (Figure~\ref{fig:scene2}). They also hear occasional voice notes from a supervisor through a hearing device and public announcements from a front-wall speaker. The user primarily focuses on the audio handbook and the supervisor's voice notes.

\textbf{Sound events and user required actions:}
In this environment, ambient noises include random crowd sounds like chatting and footsteps, and occasional sounds of people opening and closing sliding doors. 
Participants were asked to press keys 1 to 4 to identify new sentences in the audio handbook, supervisor's voice notes, knocking, and public announcements, respectively. 
The audio handbook has a one-second pause between sentences. 
Each of the four sound types occurs five times randomly during the simulation, totaling 20 sound events to be identified.

\textbf{Sound manipulations:}
\begin{itemize}
\vspace{-.3pc}
\item \proxy{Transparency Shift}: applies full noise cancellation by default and dynamically switches to half acoustic transparency during public announcements.
\item \proxy{Envelope Shift}: prioritizes the four sound types by volume: knocking, public announcements, audio handbook, and voice notes.
\item \proxy{Position Shift}: places audio handbook on the left and voice notes on the right.
\item \proxy{Time Shift}: delays virtual sounds until after the RW sounds end.
\end{itemize}

\begin{figure}[b!]
\begin{center}
\includegraphics[width=\linewidth]{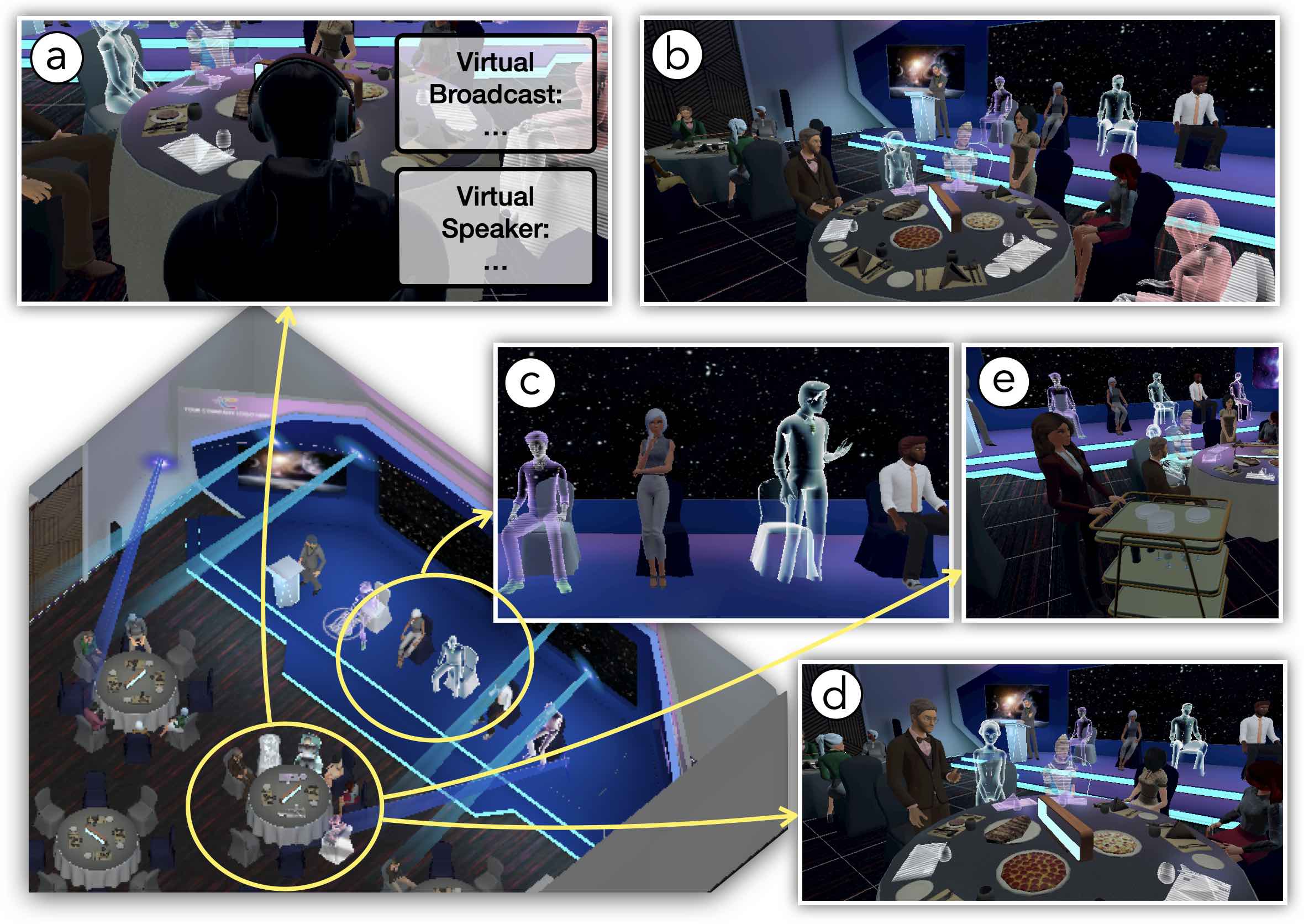}
\vspace{-2pc}
\caption{Simulated \textit{Fully-Mixed} Scenario. 
(a) The user's avatar sits at the dining table and wears headphones to consume the voice of virtual speakers and virtual broadcasts.
(b) There are physical and remote virtual speakers on the front stage, (c) where they may stand and speak up at the same time in a panel discussion, (d) similar to the physical and virtual attendees around the table. 
(e) Waitstaff sometimes comes to clean the table, generating dish clinking sounds.
}
\label{fig:scene3}
\Description{Figure 4 
A global view of the Unity scene of Fully-Mixed scenario. 
Figure 4a
A third-person view looking at the back of the person, who wears headphones sitting in front of the round table with people. There are text showing “Virtual Broadcast” and “Virtual Speaker”.
Figure 4b
A zoom out view of the reception room. There are crowds of people, including virtual and physical speakers. The virtual ones are presented by holo presence. 
Figure 4c
Some speakers sit on the chair while others stand up and talk.
Figure 4d
In the round table, there are also virtual and physical attendees, there are foods on the table and glasses of water. 
Figure 4e 
There are waitstaff coming to the table with the food cart.
}
\end{center}
\end{figure}

\subsubsection{Fully-Mixed Scenario: Attending a hybrid conference.}
\leavevmode 
\par
\textbf{Purpose:}
This setting comprises interwoven RW and VR events. We aimed to explore whether sound manipulations can harmonize both RW and VR sounds to maintain user awareness of both worlds.

\textbf{Motivation:} Our content analysis revealed several instances of simultaneous important RW and VR sounds (section \ref{need_for_RW_VR_sounds}). Also, given the prevalence of hybrid meetings in recent years and the experiences of BVI individuals in virtual settings (Section \ref{need_for_RW_VR_sounds}), we simulated this near-future scenario featuring speakers and occasional sound events from both realities.

\textbf{Scenario:}
This scenario envisions a user at a 2050 hybrid conference with attendees participating in-person or through virtual avatars (Figure~\ref{fig:scene3}), whose voices are spatially rendered based on their 3D locations. 
The user hears voices from both RW and VR, needing to discern real from virtual speakers for interaction. 
Additionally, the sound of waitstaff cleaning up, like dish-clinking, demands attention, as the user may need to make room to allow the waitstaff to pass or clear the table.
Virtual broadcasts occasionally announce events or notifications, requiring users to check their hearing devices. 
Here, the user equally pays attention to both RW and VR events.

\textbf{Sound events and user required actions:}
In this setting, ambient noises include crowds at different tables. 
There are six real-world and virtual panelists on the stage, and six attendees around the user's table. 
Virtual speakers' voices mimic an old-school phone style. 
To add complexity, each speaker overlaps with another during the panel discussion. Participants were asked to press keys 1 to 4 to identify an RW person speaking, a VR person speaking, dish-clinking, and virtual broadcasts, respectively. 
There are five instances each for table cleaning and virtual broadcasts. 
Combined with the twelve individuals speaking, there are 22 sound events in total to be identified. 
This scenario, along with the other two, is designed to last approximately 90 seconds, ensuring a balanced experience across all three scenarios.

\textbf{Sound manipulations:}
\begin{itemize}
\vspace{-.3pc}
\item \proxy{Transparency Shift}: applies half acoustic transparency.
\item \proxy{Envelope Shift}: prioritizes the four sound types using volume in the order of virtual/real people's voices, table cleaning, and virtual broadcasts.
\item \proxy{Position Shift}: places virtual broadcasts on the right.
\item \proxy{Style Shift}: applies a heavier old-school telephone effect to virtual voices.
\item \proxy{Time Shift}: delays virtual sounds until after the RW sounds end.
\item \proxy{Sound Append}: appends two separate earcons for table cleaning and virtual broadcasts.

\end{itemize}

\subsection{Technical Details on the Unity Implementation}
In each trial, the sounds were randomly scheduled and placed in different 3D locations to avoid learning effects on sound profiles across trials. 
In Unity, RW sounds were spatialized based on their 3D locations.
VR sounds were also able to be rendered spatially but mainly on either the left or right ear, or both in our study (Section \ref{walkthrough}), using commodity hearing devices.
For \proxy{Time Shift}, the known sound schedules in Unity allowed us to adjust sound characteristics (e.g., volume, pitch, and duration) to differentiate overlapping sounds or reschedule them to avoid conflict. 

For \proxy{Transparency Shift}, objects producing sound were labeled as either RW or VR, with noise cancellation effects applied only to RW objects.
We also used a headphone-blocking volume level $\eta$ to simulate the volume level reduced when noise-canceling headphones block the ears. 
By adjusting the level of transparency $\tau$ and the headphone-blocking volume level $\eta$, the volume of RW sound sources $S_{\text{new}}$ was determined by:
\begin{equation}
	\begin{aligned}
		\label{eq:ambience-volume}
    &S_{\text{new}} = S_{\text{default}} - \left( (1-\tau) \cdot S_{\text{default}}  \cdot \eta \right) \\ 
    &\tau \in [0, 1], \eta \in [0, 1]
	\end{aligned}
\end{equation}
where we set the default volume of sounds $S_{\text{default}} = 0.5$ to allow room for further manipulations (e.g., increasing volume by \proxy{Envelope Shift}), as in Unity the volume ranges from 0 to maximum 1.
For the headphone-blocking volume-reduced level $\eta$, we set $\eta=0.75$ in our implementation, calculated based on the maximum amount of noise that can be canceled by wearing active noise-cancellation (ANC) headphones (45 dB, as reported by \cite{advancedhearing}), and the average noise level in real-world environments (60 dB, as reported in \cite{normalnoise}). 
Furthermore, a high pass filter is added to all the RW sound sources and its frequency cutoff $C$ can be determined based on the transparency level $\tau$:
\begin{equation}
	\begin{aligned}
		\label{eq:high-pass-cutoff}
        C = (1-\tau) \cdot Z
	\end{aligned}
\end{equation}
where we set the baseline frequency cutoff to be $Z= 2 \text{kHz}$, since we wanted to make the half transparency in {\name} condition (when $\tau$=0.5, $C$=1kHz) matched to today's ANC technologies that can filter out signals below $1 \text{kHz}$ according to a recent report \cite{standardANC}. 
In our study, the Full Transparency condition was configured as $\tau=1$, resulting in $C=0$ kHz and volume $S_{\text{new}}=S_{\text{default}}$. For the Noise Cancellation condition, the settings were $\tau=0$, $C=2$ kHz, and volume $S_{\text{new}}=0.25 \cdot S_{\text{default}}$.

\subsection{Apparatus}
\begin{figure}[t]
\begin{center}
\includegraphics[width=\linewidth]{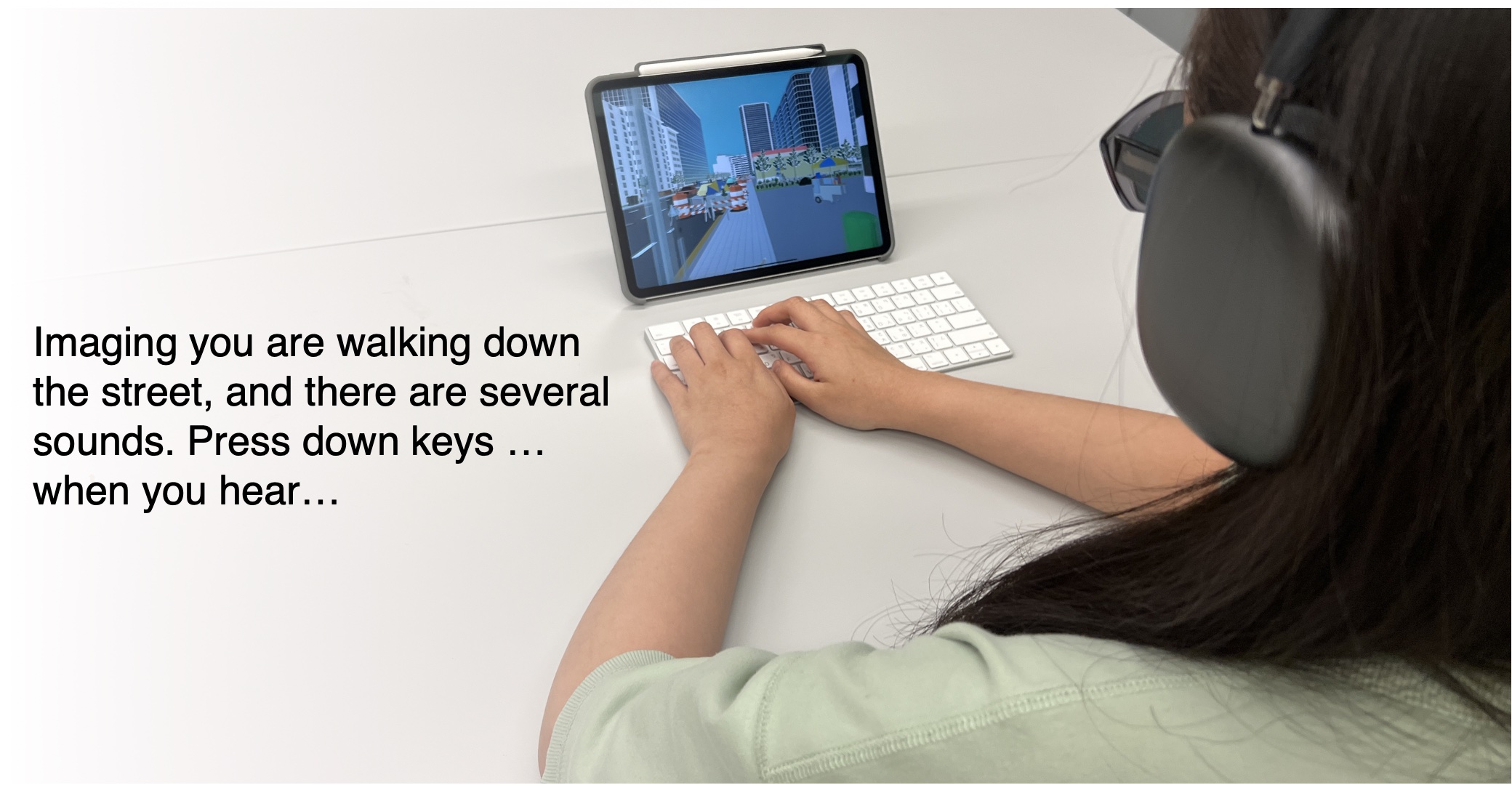}
\vspace{-1.5pc}
\caption{
In our study, participants wore headphones, engaged in pre-defined scenarios, and pressed specific keys upon hearing corresponding sounds.
}
\vspace{-1pc}
\label{fig:apparatus}
\Description{Figure 5
A user sitting in front of the table. There are a tablet and keyboard on the table. The RW-Focused scenario is shown on the tablet. There is text in the figure “Imaging you are walking down the street, and there are several sounds. Press down keys …. When you hear …”
}
\end{center}
\end{figure}

The studies were conducted in person, where participants used Apple AirPods Max provided by us (see Figure \ref{fig:apparatus}).
We allowed participants to choose whether to use the AirPods Max's active noise cancellation mode, as some BVI individuals in our pilot study found it uncomfortable.
However, we avoided the transparency mode to prevent the amplified real-world noises from distracting the participants.
The studies took place in a soundproof, enclosed space to minimize distractions from external sounds.
Participants completed the study tasks using a wireless keyboard, which, along with the AirPods Max, was connected via Bluetooth to a smartphone or tablet running our Unity application.
Users' head movements were tracked using the gyroscope data from the AirPods Max to enable spatial audio.

\begin{figure*}[ht]
    \centering
    \begin{subfigure}{0.7\linewidth}
         \centering
         \includegraphics[width=\linewidth]{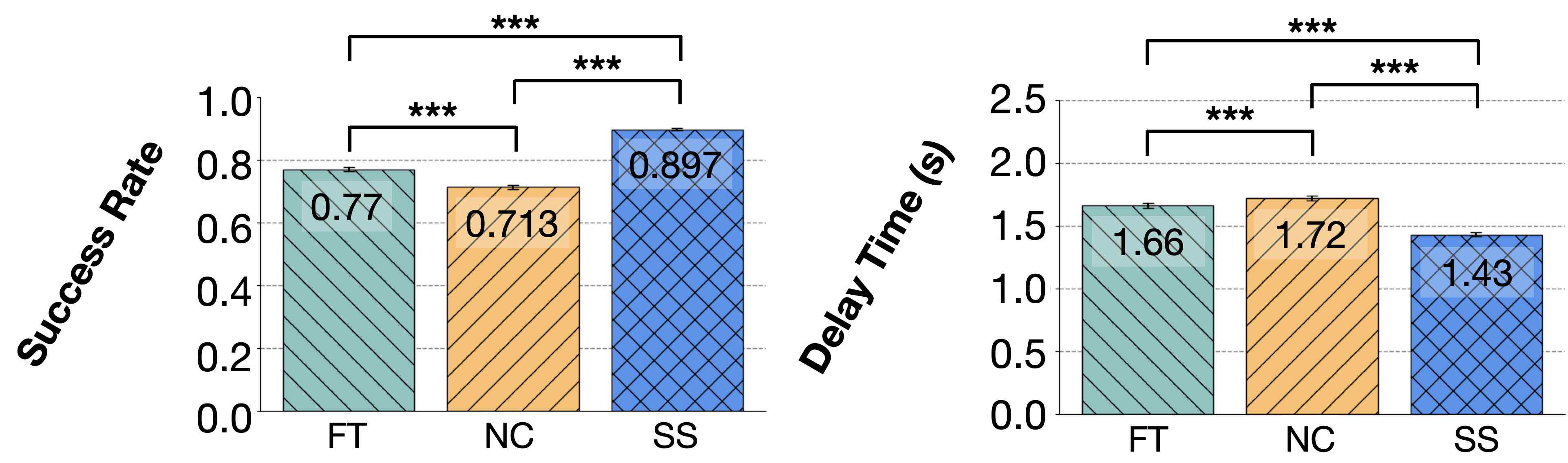}
         \begin{minipage}{1.0\linewidth}
            \vspace{-1.pc}
            \caption{Overall results based on \emph{Condition} in response to RQ1.}
         \end{minipage}
        \label{fig:conditions}
    \end{subfigure}

    \begin{subfigure}{0.7\linewidth}
        \centering
         \includegraphics[width=\linewidth]{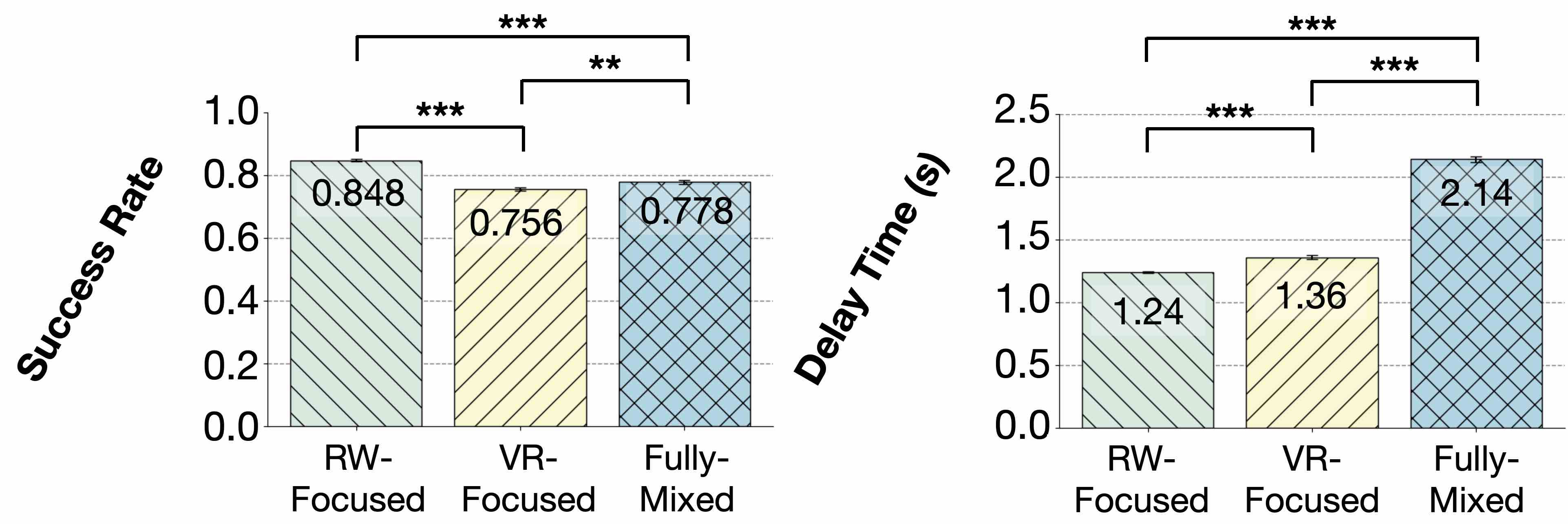}
         \begin{minipage}{1.0\linewidth}
            \vspace{-1.pc}
            \caption{Overall results based on \emph{Scenario} in response to RQ2.}
         \end{minipage}
        \label{fig:scenarios}
    \end{subfigure}
    \vspace{-1.5pc}
    \caption{Results based on \emph{Condition} (RQ1) or \emph{Scenario} (RQ2). **=$p$<=0.001. ***=$p$<0.0001.}
    \label{fig:overall}
    \Description{Figure 6a
This chart shows the results by conditions. Detailed results can be referred to Section 5.1.
Figure 6b
This chart shows the results by using the scenarios. Detailed results can be referred to Section 5.2.
}
\end{figure*}

\subsection{Tasks and Procedure}
After being welcomed and presented with our study’s informed consent and procedure, participants experienced three simulated scenarios (RW-Focused, VR-Focused, Fully-Mixed) under three conditions: Full Transparency (FT), Noise Cancellation (NC), and {\name} (SS). 
The order of scenarios and conditions was counterbalanced for the 18 participants.
Participants repeated each condition for five trials, which constituted a session.

Before each session, we briefed participants on the scenario, the four sounds to identify, and their key commands (e.g., 1,2,3,4).
They were instructed to complete each task ``as quickly and as accurately as possible'' without sacrificing accuracy for speed and vice versa.
They also had a practice session to get familiar with the individual sound events and learn the audio-key mapping. Participants could adjust the volume to their preference and take breaks anytime if needed.

After each session, we verbally described the NASA TLX form \cite{nasatlx} to our participants and obtained their responses as workload measures. 
As mentioned in section \ref{walkthrough}, there are 20 sound events in the \emph{RW-Focused} scenario, 20 for the \emph{VR-Focused} scenario, and 22 for the \emph{Fully-Mixed} scenario.
Therefore, for each participant, we collected data from 62 audio events $\times$ 3 conditions $\times$ 5 trials, resulting in 930 tasks. This amounts to 930 tasks $\times$ 18 individuals = 16,740 tasks.
However, some data points were excluded due to technical issues, resulting in 16,700 effective tasks.
The study, approved by the IRB, compensated each participant with 40 USD. It took about 2 hours, with 80 minutes for task completion and 40 minutes for NASA-TLX responses and other follow-ups.

\subsection{Dependent Measures and Data Analysis}
For each trial, we recorded task data, focusing on two primary dependent measures: 
\textit{(i) Success Rate}, calculated as the ratio of correct key presses to total sound events, and 
\textit{(ii) Delay Time}, considering only correct key presses and measuring the time from the event's onset to the correct key press. 
We conducted a mixed-methods analysis. 
We built two separate mixed-effect linear regression models \cite{lindstrom1988newton} to examine the dependent variables \emph{Success Rate} and \emph{Delay Time}, with fixed effects \emph{Condition} and \emph{Scenario}, and their interaction \emph{Scenario $\times$ Condition},  
taking participant ID as a random intercept. 
We also transcribed our participants’ feedback from each session for further analysis.

\begin{figure*}
     \centering
     
     \begin{subfigure}[t]{0.7\linewidth}
         \centering
         \includegraphics[width=\linewidth]{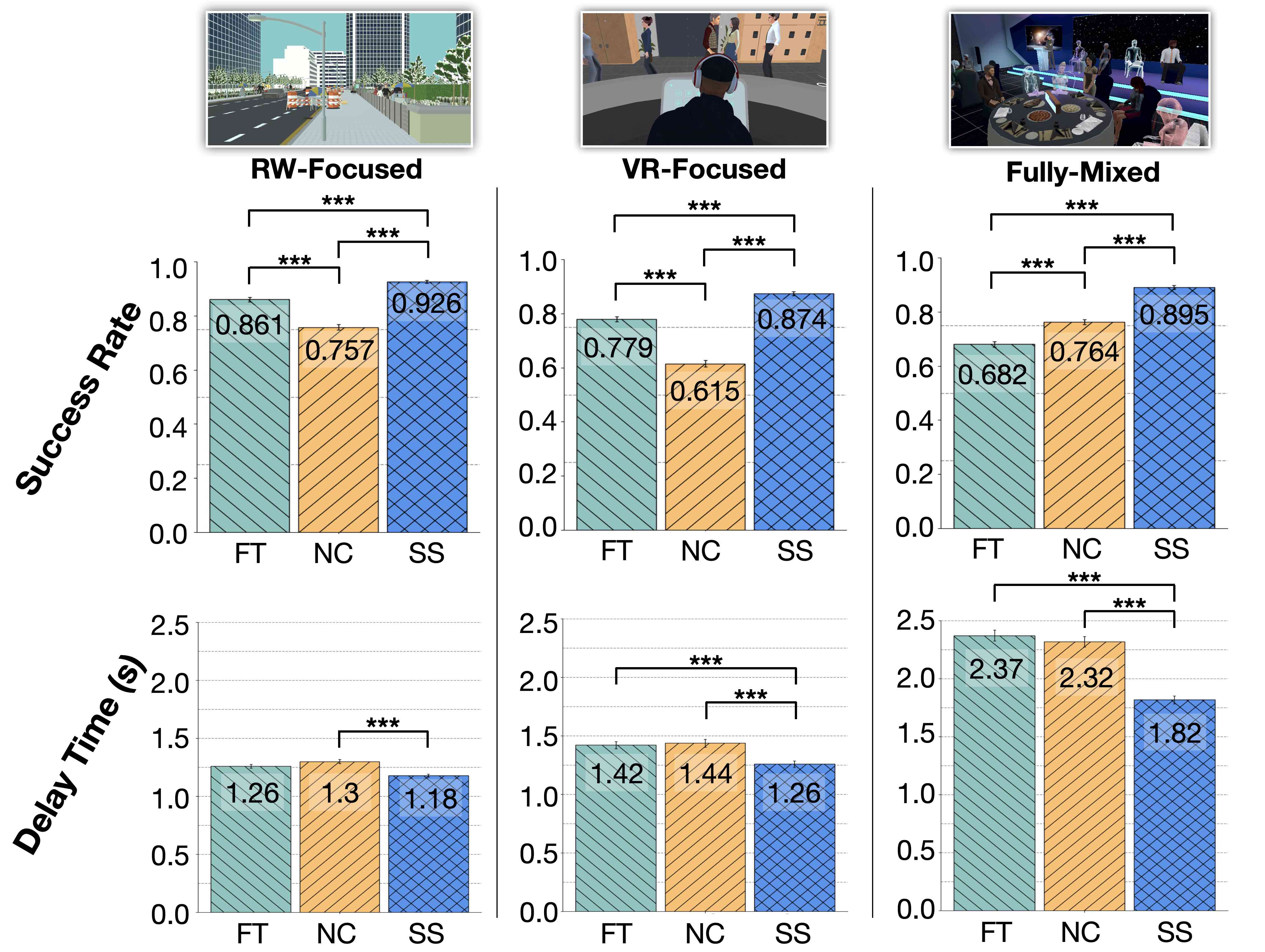}
         \begin{minipage}{0.9\linewidth}
            \vspace{-1.pc}
            \caption{}
         \end{minipage}
         
         \label{fig:results}
     \end{subfigure}
     \hfill
     \begin{subfigure}[t]{0.295\linewidth}
         \centering
         \includegraphics[width=0.82\linewidth]
         {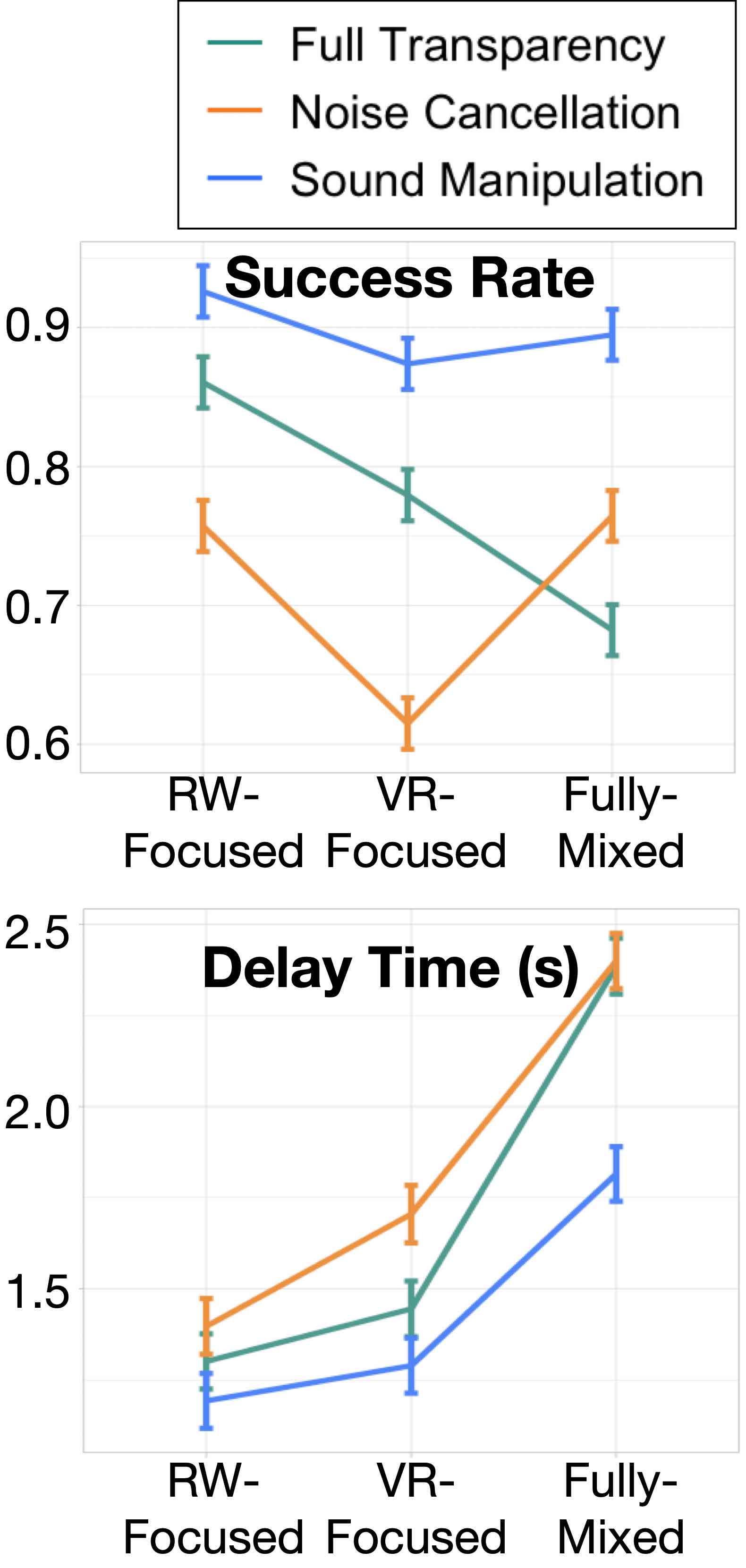}
         \begin{minipage}{1.05\linewidth}
            \vspace{-1.pc}
            \caption{}
         \end{minipage}
         \label{fig:interaction}
     \end{subfigure}
     \vspace{-2.pc}
     \caption{Overall results for RQ3 based on \emph{Condition} and \emph{Scenario}, including (a) \emph{Success Rate} (top, the higher the better) and \emph{Delay Time} (bottom, the lower the better) with error bars showing 95\% confidence intervals. ***=$p$<0.0001. 
     (b) Interaction effects on \emph{Success Rate} and \emph{Delay Time} between \emph{Condition} and \emph{Scenario}.
     }
     \label{fig:overall_results}
     \Description{Figure 7 This chart shows the results by examining both conditions and scenarios. Detailed results can be referred to Section 5.3. (a) \emph{Success Rate} (top, the higher the better) and \emph{Delay Time} (bottom, the lower the better) with error bars showing 95\% confidence intervals. ***=$p$<0.0001. 
     (b) Interaction effects on \emph{Success Rate} and \emph{Delay Time} between \emph{Condition} and \emph{Scenario}.
    }
\end{figure*}

\subsection{Results}
This section reports our study's quantitative and qualitative results to answer each of our research questions.

\subsubsection{RQ1: How do sound manipulations affect participants' performance compared to full transparency and noise cancellation?} \label{condition_results}
\textbf{With \textit{{\name}}, participants achieved a significantly higher success rate and lower time delay in identifying sounds than FT and NC, but sound manipulations may confuse users' interpretation of the sound content.
}

Overall, we found that \emph{Condition} had a significant main effect on both \emph{Success Rate} ($F$(2,16700)=327.58, $p$<0.0001) and \emph{Delay Time} ($F$(2,13242)=88.02, $p$<0.0001) in the three scenarios (Figure \ref{fig:overall}a).
Post-hoc Tukey's pairwise test revealed that SS (M=0.897) resulted in a significantly higher \emph{Success Rate} than FT (M=0.770) and NC (M=0.713), and FT was also significantly higher than NC, $p$<0.0001; for \emph{Delay Time}, SS (M=1.43s) was significantly lower than FT (M=1.66s) and NC (M=1.72s), and FT was significantly lower than NC, $p$<0.0001. 
It was expected that SS would increase sound awareness in both RW and VR, and all participants commended the benefits of {\name}: 
\QUOTE{SS achieves proper volume without the problem of sounds being too noisy or mixed up} (B16, \emph{RW-Focused}), 
\QUOTE{SS is quite authentic and echos with my custom ... I like to wear headphones on the right ear only when walking} (B12, \emph{RW-Focused}), 
\QUOTE{I like SS as I also used to listen to certain things on one side} (B8, \emph{VR-Focused}), 
and \QUOTE{I like it since it makes sounds very apparent, like notification of table cleaning, and the real people's voices are also different from broadcast} (B3, \emph{Fully-Mixed}).
Also, the reason that FT led to better performance than NC was due to the nature of NC that blocks out RW sounds, as remarked by most participants and evidenced by the \emph{Success Rate} of RW (M=0.693) and VR (M=0.895) sounds, and \emph{Delay Time} of RW (M=1.81s) and VR (M=1.51s) sounds. 
Also, FT was preferable as it retained RW awareness and was more familiar to BVI participants, echoing the results in section \ref{retain_RW_awareness} that BVI people emphasized the importance of retaining RW awareness when surveying different headphones.

However, participants also commented that {\name} may confuse their perception and interpretation of RW sounds, such as 
\QUOTE{... the post-processed gentle drilling made me hard to tell the scale of the construction sites} (B10, \emph{RW-Focused}), 
\QUOTE{It is weird ... knocking sounds are far away from me} (B8,\emph{VR-Focused}), and 
\QUOTE{The robot-like filter is distinguishable but distorts the content ... I need to pay a lot of attention} (B12, \emph{Fully-Mixed}). 
This highlights a trade-off between enhancing sound awareness and preserving the fidelity of certain sound characteristics when manipulating sounds.

\subsubsection{RQ2: How do the different scenarios with varying emphases on reality and virtuality affect participants’ performance?}\label{scenario_results}
\textbf{Participants overall achieved a higher success rate and lower time delay in RW-Focused than VR-Focused and Fully-Mixed scenarios due to their familiarity.}

We found that \emph{Scenario} had a significant main effect on both \emph{Success Rate} ($F$(2,16700)=83.22, $p$<0.0001) and \emph{Time Delay} ($F$(2,13242)=810.83, $p$<0.0001). 
Post-hoc Tukey's pairwise test revealed that participants had significantly higher \emph{Success Rate} in \emph{RW-Focused} (M=0.848) than in \emph{VR-Focused} (M=0.756) and \emph{Fully-Mixed} scenarios (M=0.778), $p$<0.0001 (Figure \ref{fig:overall}b). 
Furthermore, \emph{Fully-Mixed} was significantly higher than \emph{VR-Focused}, $p$=0.001.
Participants also had significantly lower \emph{Delay Time} in \emph{RW-Focused} (M=1.24s) than in \emph{VR-Focused} (M=1.36s) and \emph{Fully-Mixed} (M=2.14s) scenarios, $p$<0.0001; \emph{VR-Focused} was significantly lower than \emph{Fully-Mixed}, $p$<0.0001.

Overall, results suggested that the \emph{RW-Focused} scenario yielded better performance, possibly due to participants' familiarity with the navigation scenario and the RW sounds: \QUOTE{This is how I get to my home, but mine is more complicated as typically sounds are not evenly distributed and consistent} (B10), and \QUOTE{[FT] is helpful as it's very like the real situation ... made me easily engage in} (B15). 
Conversely, performance in the \emph{VR-Focused} and \emph{Fully-Mixed} scenarios was less effective, possibly due to their unfamiliarity with the audio content.
Specifically, some participants mentioned the audio handbook's sentences in the \emph{VR-Focused} scenario were new and too long to focus on, while the overlapping conversations in the \emph{Fully-Mixed} scenario made the content hard to observe, not to mention further distinguishing its source from RW to VR, as argued by B10: \QUOTE{I don't think the real or virtual person talking matters ... I cannot even hear clearly on the content when they are totally mixed.}
These results were expected due to our futuristic and complex design of \emph{Fully-Mixed} scenario, where participants' unfamiliarity with the content and scenario affected their performance.

\begin{figure*}[t]
\begin{center}
\vspace{-1pc}
\includegraphics[width=1.0\linewidth]{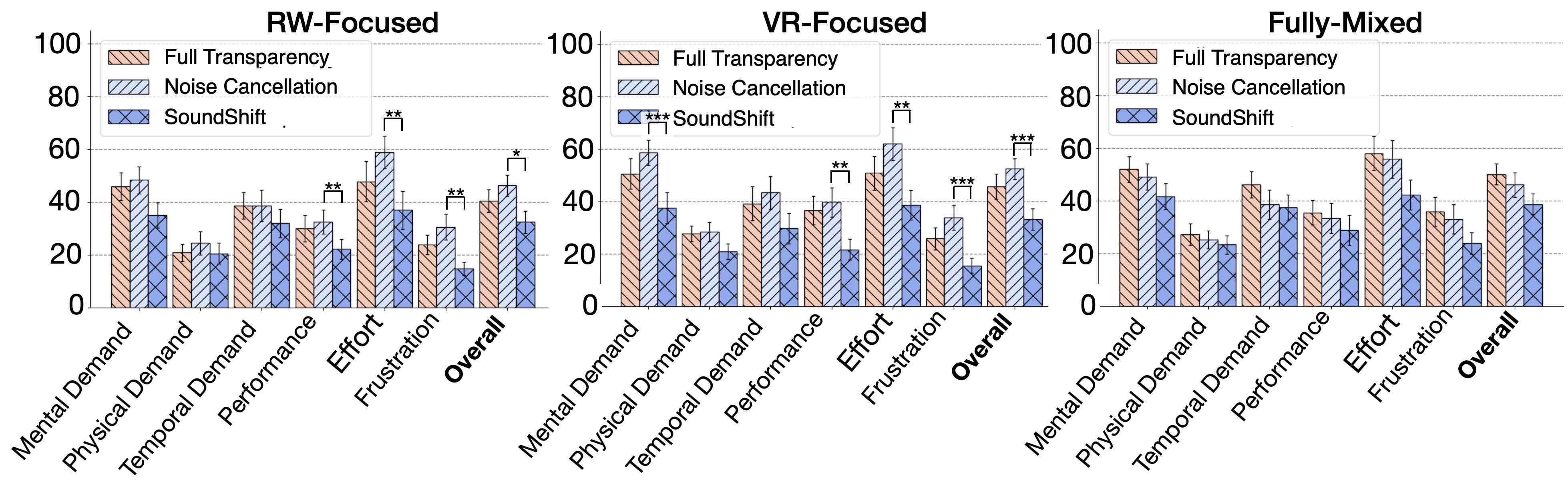}
\vspace{-2pc}
\caption{Results of NASA-TLX (RQ4) with error bars showing 95\% confidence intervals. ***=$p$<0.0001, **=$p$<0.001 and *=$p$<0.01.
}
\label{fig:NASA}
\Description{Figure 8
This chart shows NASA-TLX results for RQ4. Detailed results can be referred to in Section 5.4.
}
\end{center}
\end{figure*}

\subsubsection{RQ3: How do the conditions affect participants’ performance differently across the scenarios?}\label{interaction_condition_scene_result}
\textbf{In all scenarios, SS was more effective and preferable than FT and NC, but still, many participants had varied preferences on sound awareness and found the sound clarity and naturalness of FT and the quietness of NC useful.}

We also found a significant interaction effect (Figure \ref{fig:overall_results}b) between \emph{Condition} and \emph{Scenario} for both \emph{Success Rate} ($F$(4,16700)=56.7, $p$<0.0001) and \emph{Delay Time} ($F$(4,13242)=23.96, $p$<0.0001). 
Specifically, in the \emph{RW-Focused} scenario, post-hoc Tukey's pairwise test revealed that participants had significantly higher \emph{Success Rate} with SS (M=0.926) than FT (M=0.861) and NC (M=0.757), and FT than NC, $p$<0.0001 (Figure \ref{fig:overall_results}a); they also had significantly lower \emph{Delay Time} with SS (M=1.18s) than NC (M=1.30s), $p$<0.0001.
This was evidenced by most participants (N=14) that SS was preferable and provided a clear sound awareness of both RW and VR sounds than FT and NC, as B18 stated \QUOTE{I love the first one [SS], because the ambient noise in the second one [FT] is excessively noisy, and the construction sound in the third one [NC] is almost inaudible, which is very dangerous.}
On the other hand, four participants liked FT the most as \QUOTE{... the sound of the white cane is more distinctive. The drilling is more vivid, which makes me more aware of the situation on the road ... It [FT] is helpful because the sound is pretty spatial, I am able to perceive the composition of the environment} (B14).
Participants also indicated that while NC made instructions clearer, it also lost information from RW, which was unfavorable in navigation scenario.

In the \emph{VR-Focused} scenario, post-hoc Tukey's pairwise test revealed that participants achieved a significantly higher \emph{Success Rate} with SS (M=0.874) than FT (M=0.779) and NC (M=0.615), and FT than NC, $p$<0.0001 (Figure \ref{fig:overall_results}a); they also had significantly lower \emph{Delay Time} with SS (M=1.26s) than FT (M=1.42s, $p$=0.0065) and NC (M=1.44s, $p$<0.0001). 
Fourteen participants preferred and performed better in SS for its distinct audio channels, with the audio handbook on the left and the supervisor's voice note on the right:
\QUOTE{All details can be observed. The separate audio channel design is quite good, especially in such a static environment ... it allowed me to distinguish different sounds from different channels ... ease my burden} (B14).
Another three participants liked SS and FT equally as FT retained the fidelity of sounds, as remarked by B8: \QUOTE{Both SS and FT have aspects I like. FT makes it a real-life scenario, while the information in SS is very clear.}
One participant (B13) preferred NC for its quieter environment due to his high sensitivity to sounds.

In the \emph{Fully-Mixed} scenario, post-hoc Tukey's pairwise test revealed that participants performed significantly better with SS (M=0.891) than FT (M=0.68) and NC (M=0.762), and NC than FT, $p$<0.0001 (Figure \ref{fig:overall_results}a); they also had significantly lower \emph{Delay Time} with SS (M=1.82s) than FT (M=2.37s, $p$<0.0001) and NC (M=2.32s, $p$<0.0001). 
Many participants (N=11) preferred SS for its sound clarity, as commented by B15: \QUOTE{It's helpful and allows me to easily differentiate between broadcasts and human voices. Also, the sounds don't seem to overlap as much.} 
However, some participants felt more pressured and overwhelmed in SS and preferred FT (N=5), as stated by B10: \QUOTE{I liked FT which has natural voices, and it has no earcons that might drown out human voices}, while others preferred NC (N=2), like B12: \QUOTE{The sound in this condition [NC] is clear and comes gradually, though it doesn't make every sound very clear. The second one [SS] makes sounds too clear ... It might cause me to overlook other important sounds.}
In this complex scenario, participants' preferences varied, and the enhanced clarity of specific sounds in SS raised concerns about potentially missing other crucial audio information.

\subsubsection{RQ4: How do sound manipulations affect participants' cognitive load compared to full transparency and noise cancellation?}\label{nasa_results}
\textbf{\textit{{\name}} overall reduced the cognitive load.}

We found a significant main effect of \emph{Condition} on the overall workload in NASA-TLX ($F$(2,136)=17.01, $p$<0.0001).
Post-hoc Tukey’s pairwise test revealed that SS (M=34.8) resulted in a significantly lower workload than both FT (M=45.4) and NC (M=48.2), $p$<0.0001, but there was no significant difference between FT and NC ($p$=0.4637). 
The reason behind this result might be explained by the previous findings, where SS increased sound awareness via different manipulations; on the other hand, both FT and NC resulted in similar cognitive load, likely due to their inherent limitations, such as the multiple overlapping sounds in FT and the nearly-blocked RW sounds in NC. 
Interestingly, we found no significant differences between SS and the other two conditions for the overall cognitive load in the \emph{Fully-Mixed} scenario.

This might be due to the result in section \ref{interaction_condition_scene_result} that the stylized voices in SS sometimes made content unclear, or the enhanced clarity of specific sounds in SS might overshadow other crucial audio information.

\subsubsection{RQ5: How do participants describe their experiences and ways to further customize their soundscape for each scenario?}
\label{customization_results}
\textbf{Participants suggested retaining natural sounds and customizing sound manipulations based on the content and context of sounds.} 

Most participants suggested dynamically increasing the volume of certain sounds or reducing background noises to improve identification performance. 
They also noted using different filters for certain sounds for better comfort. 
Several participants proposed various customizations beyond the sound manipulations used in our study.

First, most participants favored the natural presentation of RW sounds to maintain their perception of the real world, (Section \ref{condition_results}).
Yet, some participants desired more intelligent sound manipulations based on the RW context. 
For instance, B1 said in the \emph{RW-Focused} scenario
\QUOTE{I would only mute the voices of strangers but not all people as I want to hear the voices of someone I know}, and B10 also mentioned \QUOTE{I think earcon should be played only when the construction site is large enough, as typically, a small scale of construction would not be so noisy and discomforting.} 
Similarly, most participants also preferred keeping the voices of virtual people in the \emph{Fully-Mixed} scenario unaltered, not imposing another sound filter on them. 
Instead, they may use other methods, such as sound localization or discerning voice nuances, to distinguish between RW and virtual reality (VR) sources.

Participants also proposed manipulating sounds based on the audio content, where sound characteristics can adaptively change based on the urgency and importance of the content.
For instance, B15 stated in the \emph{VR-Focused} scenario that \QUOTE{I would lower down the public announcements if the content is not relevant to me}, and some participants also mentioned the volume of the supervisor's voice notes should be proportional to the urgency of the content. 
Furthermore, B3 mentioned managing the voice font for the virtual broadcast based on the content \QUOTE{I would change it to a cuter voice such as characters in the game or anime ... however, serious content can be presented in a deep voice.}

In sum, besides our proposed sound manipulations, participants also suggested that sound manipulations could be grounded on context and content to better manage their attention.

\section{Example Applications with {\name}}\label{application}
Based on insights from our studies, we developed three proof-of-concept prototypes to showcase the practicality and generalizability of {\name} manipulations in different MR experiences.
Though BVI people did not evaluate these applications, we hoped these applications stimulate further discussions and advancements in sound manipulation for future mixed-reality applications to promote accessibility.
Note that these prototypes are functional. We highlight user interactions rather than technical specifics in this section. Please refer to our Video Figure for demonstrations.

\begin{figure}[h]
\begin{center}
\includegraphics[width=\linewidth]{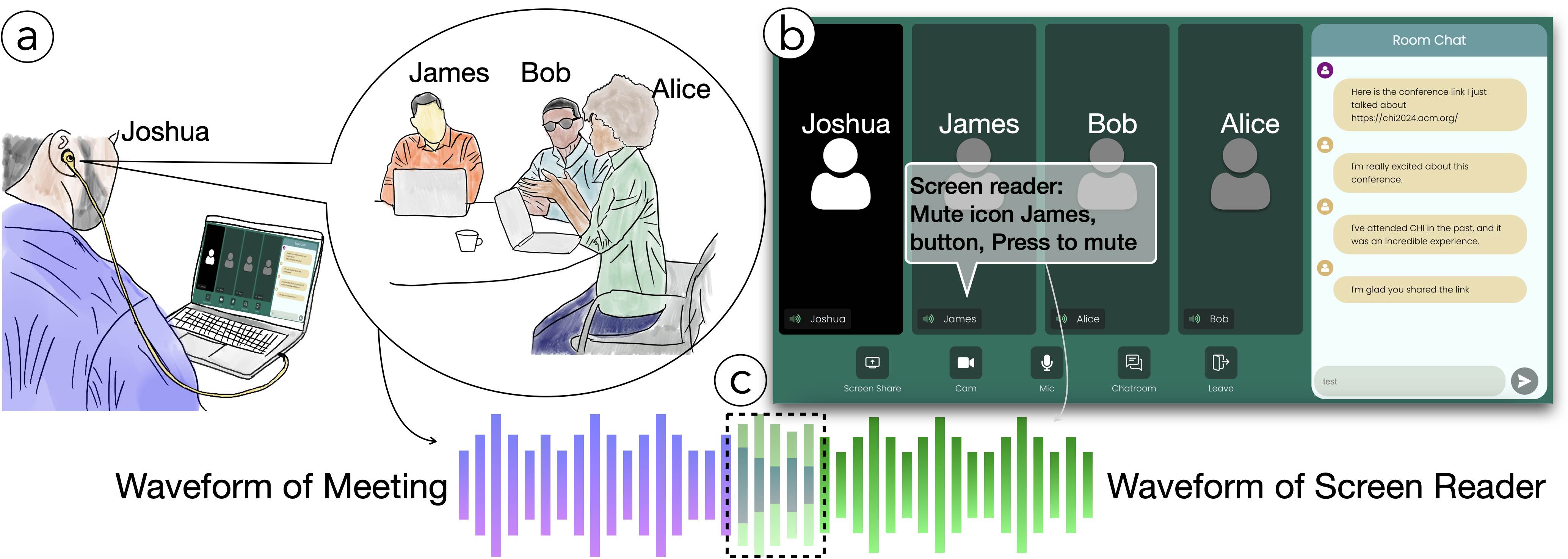}
\vspace{-1.8pc}
\caption{A prototype of an accessible online meeting application and demonstrated waveforms showing how sounds are manipulated. 
(a) The voices of the three people are assigned to the left, front, and right spatial locations around the BVI user, same as (b) the layout of the attendee panel.
(c) When using a screen reader to navigate, the meeting volume is decreased to accentuate the audio feedback of the screen reader.
}
\vspace{0pc}
\label{fig:meeting}
\Description{Figure 9
Figure 9a
Joshua is a person who attends a virtual meeting with James, Bob, and Alice. This figure conceptually showing the the spatial audio enabling the atmosphere of physical meeting. 
Figure 9b 
The interface of the web application. There are attendee panels on the interface and some basic widgets.
Figure 9c
Waveforms show that the volume of screen reader will be louder than meeting when they are in conflict
}
\end{center}
\end{figure}

\subsection{Accessible Online Meeting Application}
In response to difficulties highlighted in section \ref{need_for_RW_VR_sounds} about using screen readers during virtual meetings, we developed an audio-adaptive online meeting web application.
This app resolves conflicts between screen reader and meeting audio using \proxy{Time Shift}, seamlessly blending these sounds by adjusting their auditory characteristics.
For example, it increases the screen reader's volume when someone else speaks but allows users to prioritize meeting audio if it holds higher priority. 
To address the lack of spatial awareness common in virtual meetings, \proxy{Position Shift} arranges attendees' voices across the left-right audio spectrum, corresponding to their positions on the attendee panel. 
This feature lets users easily map the voice's location to the attendee's location on the interface for further interaction (e.g., sending private messages).

\subsection{Content-Aware MR Image Exploration}
\begin{figure}[t]
\begin{center}
\includegraphics[width=\linewidth]{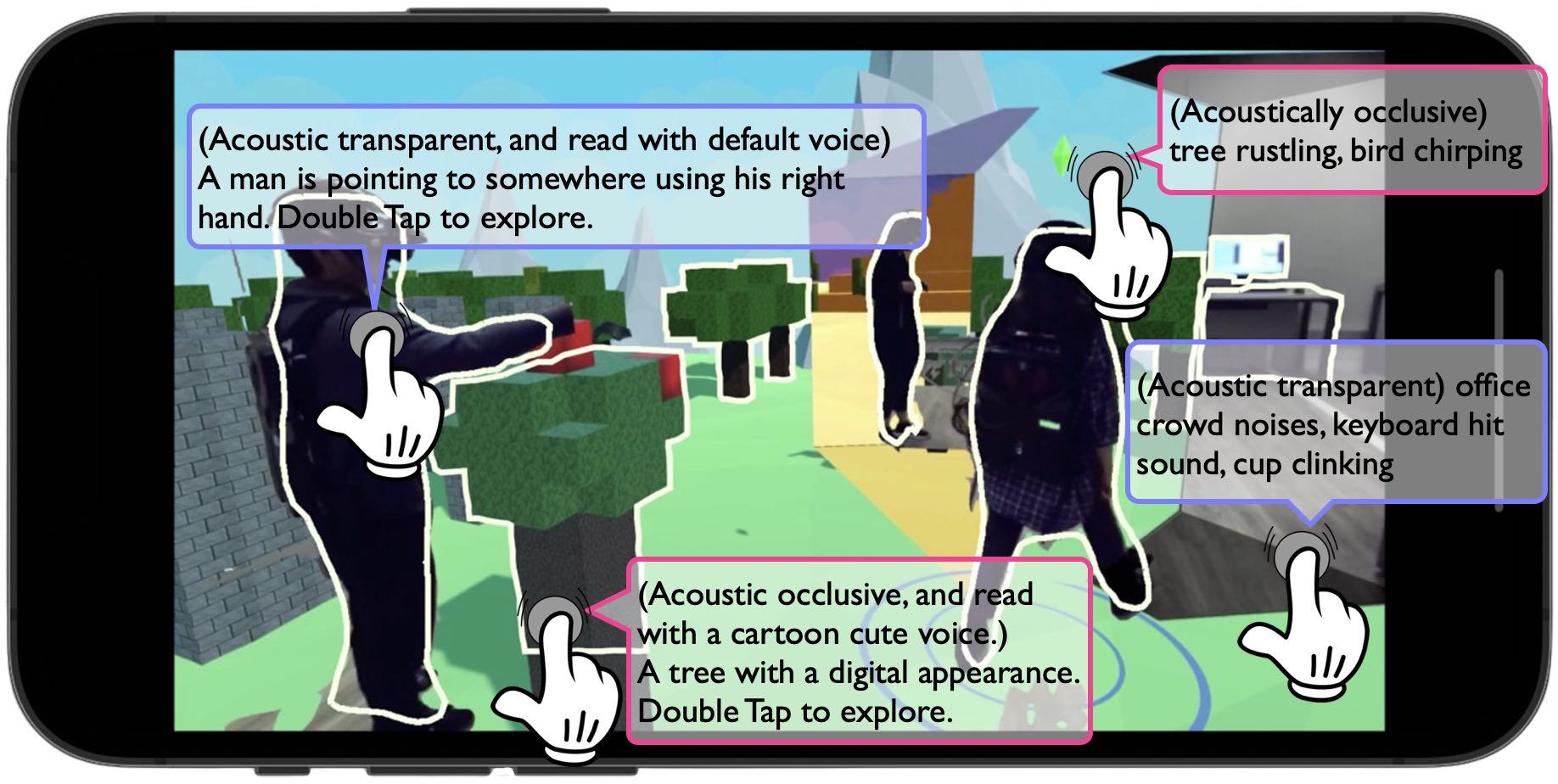}
\vspace{-1.8pc}
\caption{
Using ImageExplorer \cite{imageexplorer} with sound manipulations. The example image was provided by Wang et al. \cite{sliceoflight}, which includes a combination of real and virtual content. 
The voice corresponding to the touch locations is spatialized from left to right.
The voice of VR content is read with a cartoon-ish voice to fit the digital cartoon-style environment, while the voice of RW content remains as default.
Noise cancellation mode is turned on when exploring VR content, and shifts to transparency mode when exploring RW content. 
} 
\label{fig:imageexplorer}
\Description{Figure 10
A smartphone showing a mixed reality image. The image contains several physical people wearing the head-mounted display. There are a open office space on the right of the image. Otherwise, people are seen as in the virtual park, which has digital cube-based trees and sky view. 
}
\vspace{-1pc}
\end{center}
\end{figure}

Mixed-reality images are increasingly pervasive, such as in news or research articles, which entail rich visual information.
However, the spatial information is not adapted from the visual to the audio domain.
In response, our image touch exploration system built on \cite{imageexplorer} spatializes the screen reader's voice feedback by mapping the touch location on the image to the sound feedback's spatial location (Figure \ref{fig:imageexplorer}) for enhancing spatial understanding.
Moreover, MR images, which combine digital and physical visuals easily distinguished visually but inaccessible to BVI people.
To address this, during image exploration, \proxy{Transparency Shift} renders acoustic transparency for real-world (RW) content and occlusive opacity for virtual reality (VR) content, such as crowd noises in the office for RW content and bird chirping for VR content (Figure \ref{fig:imageexplorer}), which signals the reality and virtuality of content without additional text descriptions.
Moreover, \proxy{Style Shift} alters the voice of description to a cartoonish tone to match the digital world's style and enhance immersion.

\subsection{Context-Aware Outdoor Navigation}
\begin{figure}[t]
\begin{center}
\includegraphics[width=\linewidth]{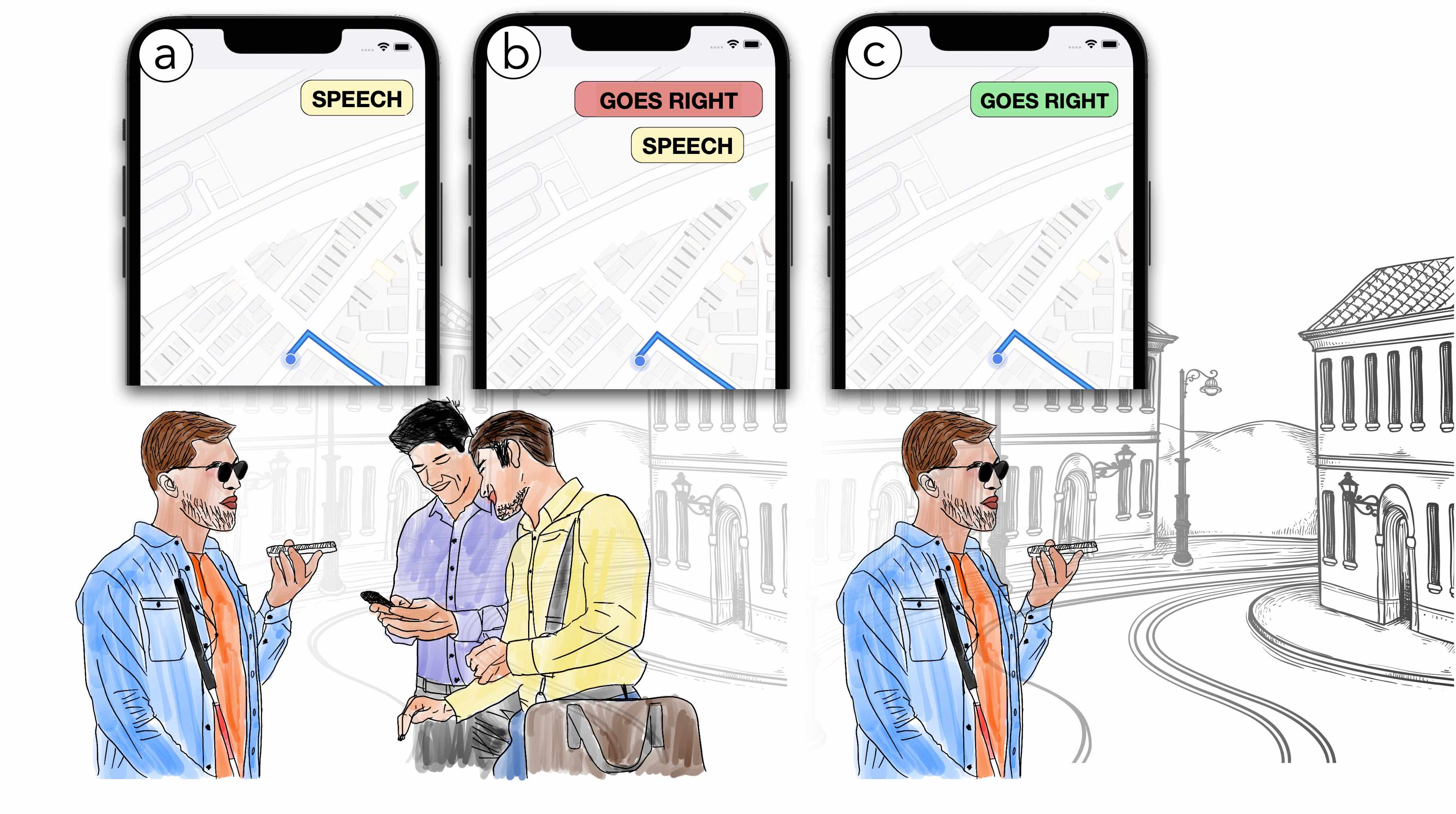}
\vspace{-2.2pc}
\caption{
Illustration of our context-aware navigation app.
(a) The crowd passing by the BVI person generates sounds that are detected by the app, and at the same time, (b) an audio direction is about to happen but will not be played due to its conflict with crowd sounds.
(c) After the crowd leaves, the audio direction is played.
} 
\label{fig:navigation}
\Description{Figure 11a
A blind person is using a smartphone, and two people are passing by and talking to each other. There is text “Speech” showing on the top right corner of the phone screen, suggesting that the speech of the two people is detected.
Figure 11b
The phone screen shows the audio direction “Go right” in the top right corner with the red background, suggesting it is still pending
Figure 11c
After the two people leave, the audio direction “Go right” becomes green, suggesting that it is played to the user. 
}
\vspace{-1pc}
\end{center}
\end{figure}

Navigation is indispensable to BVI people, and many commercial apps \cite{soundscape,blindsquare} have been developed to facilitate BVI people's independence and autonomy. 
However, occasional events may distract users from navigation.
We developed a mobile navigation application that analyzes both real-world and virtual sounds to identify opportune moments for delivering audio directions.
When a certain RW event is happening and detected, the audio directions will be delayed by \proxy{Time Shift} until the RW event ends (Figure \ref{fig:navigation}). 
If a certain RW sound is detected but overshadowed by the audio directions, this app can provide post-hoc verbal descriptions to notify users of the detected sound.

\section{Discussion and Future Work}
We have contributed the concept of \textit{{\name}} with six sound manipulators to increase MR sound awareness for BVI people, a user study to prove the effectiveness of sound manipulations on different MR scenarios, and three example applications to demonstrate its feasibility for practical use. 
In this section, we discuss our work's limitations and potential improvements, implications for future MR soundscape design, and the generalizability of our findings to broader communities.

\subsection{From Simulation to Practical Applications}
In our work, we created Unity simulations for real-time manipulation of sound characteristics and developed three proof-of-concept applications to illustrate the generalizability and practicality of {\name}.
We encountered several challenges; for instance, the audio streams in most existing platforms were at the operation system level, which made them hard to access for application-level purposes.
Furthermore, implementing these sound manipulations in the real world necessitates several requirements, 
such as supporting the recognition of diverse categories of sounds,
extracting sounds of interest from multiple overlapping ones, 
and rendering the extracted sounds in their original or higher quality. 
And all of these components should be achieved in real time to ensure a seamless user experience, which is still challenging in sound research.
These are why our three example applications mostly revolve around manipulating virtual audio.

In recent years, researchers have attempted to tackle these challenges. 
For instance, Ubicoustics \cite{ubicoustics} supports the recognition of many sound activities through commercially available microphones, Jain et al. \cite{protosound} developed ProtoSound for people to customize their sound recognition model, and Veluri et al. \cite{veluri2023real} approached real-time target sound extraction. 
Also, to increase the accuracy of sound identification or extraction, we could approach the users' sound context with other sensing modalities. 
For example, in the \emph{RW-Focused} scenario, one's smartwatch could provide gyroscope data to detect white cane taps. 
Additionally, everyday objects often produce sounds linked with visual elements, detectable through cameras, such as playing an instrument \cite{tian2021cyclic}.
These approaches can provide a preliminary context for the targeted sound identification or extraction.
Despite a long way ahead in real-time manipulation of overarching sounds, recent research endeavors have illuminated promises to apply {\name} to practical applications.

\subsection{Towards Sound-Aware Description Manipulations}
As the introduction outlines, incorporating visual descriptions is another essential auditory element to facilitate MR experiences for BVI people. 
Unlike diegetic sound effects (e.g., knocking, drilling, dog barking), descriptions provide specific information and semantic meanings crucial for BVI users.
In section \ref{customization_results}, participants suggested making virtual broadcasts or voice notes discernible amidst other sounds only if the content is deemed critical, as a strategy to mitigate information overload.
It is thus promising to adjust the description content or provide opportune sentence breaks to prevent it from overlapping with other MR auditory elements. 
For instance, the system can offer comprehensive descriptions during quieter moments, while providing succinct yet informative descriptions when time is limited, to create a harmonized user experience.
This is similar to creating audio descriptions with dynamic time constraints in a video, such as Rescribe \cite{pavel2020rescribe} to shorten the descriptions by removing less important words. 
Overall, sound manipulation in MR environments should extend beyond adjusting sound characteristics, temporal, and spatial aspects. It should also consider the content and audio context, making it more intelligent, adaptive, and user-centric.

\subsection{Customizable Sound Manipulations}
Though each manipulator in our Unity implementation can manipulate sounds in real-time (e.g., overlap of several sounds) to provide a seamless MR soundscape, we did not consider the varied preferences of our participants.
As described in section \ref{customization_results}, participants proposed several ideas on manipulating sounds based on their current context and content of audio information (e.g., sounds, text descriptions), which also echos with our formative results in section \ref{online_forum_augment_sound} that BVI people desired to augment or customize the sound library of existing applications.
It is thus promising to integrate methods that gauge the importance of audio information, such as content analysis or sentiment analysis. 
Based on the gauged importance, users could personalize sound representation (e.g., earcons, descriptions, diegetic sounds) and presentation (e.g., high/low volume, sound locations) to minimize distraction.
Additionally, participant preferences for sound manipulation, influenced by personal experiences and memories, highlight the need for adaptable soundscapes;
for instance, the sound of the sliding door in the \emph{VR-Focused} scenario made B1 recall her bad memory of consuming videos about insects, and she wanted to disable it selectively. 
Research on visual descriptions also emphasized individual customization over a one-size-fits-all approach \cite{kreiss2022context, stangl21assets, lopez2018audio, lopez2022seeing}.
Echoing this notion, it is, therefore, worth exploring customizable sound properties and how to design end-user interfaces to enable user customization.

\subsection{Generalizing Results to Broader Groups with Different Sensory Modalities}
Though our content analysis from online posts gave us broad insights into different scenarios and needs on the everyday consumption of complex sounds, it lacked depth in understanding the full context behind the posts. Future research could dive deeper into relevant topics, such as understanding diverse contexts of consuming complex sounds, the sound technologies BVI people currently use to help with everyday tasks, or their strategies for handling sounds in more futuristic and complex mixed-reality scenarios.

Furthermore, the simulations in our main study may not fully capture the multisensory experience that BVI individuals rely on, such as a combination of haptic feedback, smell, and other contextual cues, to enhance their real-world awareness. 
That was also why several participants in the \emph{RW-Focused} scenario emphasized the importance of haptic feedback from the white cane besides its auditory feedback. 
Nonetheless, while other sensory feedback can augment and address the limitations of sound, they are all of significance and cannot replace one another. 
Also, none of our participants consumed visual feedback despite some having residual visual ability. Instead, they used hearing only to perform the tasks.
Future work could investigate how people strategize their sensory modalities for balancing cognitive load and performance in mixed reality, which may enable manipulators across modalities (e.g., split conflicting audio information into visual and audio ones) to create more sensory-adaptive and accessible mixed reality experiences.

\section{Conclusion}
We have presented the concept of {\name} to make MR sound awareness accessible for BVI people, through six sound manipulators derived from our content analysis on BVI forums, including \proxy{Transparency Shift}, \proxy{Envelope Shift}, \proxy{Position Shift}, \proxy{Style Shift}, \proxy{Time Shift}, and \proxy{Sound Append}.
We instantiated the six sound manipulators and three simulated scenarios across the Reality-Virtuality continuum in Unity.
We then conducted a user study with eighteen BVI people and found empirical evidence that the six sound manipulations significantly enhanced users' sound awareness and reduced cognitive load.
We also found varied preferences and comments across participants on manipulating sounds, which spurred several discussions and promises of future work.
Finally, we implemented three proof-of-concept applications to demonstrate the generalizability and practicality of {\name}, including an accessible online meeting app, an immersive image understanding system, and a context-aware navigation app.

\begin{acks}
We thank our anonymous reviewers for their suggestions and all the participants in our study.
\end{acks}

\bibliographystyle{ACM-Reference-Format}
\bibliography{sound}


\begin{thebibliography}{88}

%%% the Web address.

\ifx \showCODEN    \undefined \def \showCODEN     #1{\unskip}     \fi
\ifx \showDOI      \undefined \def \showDOI       #1{#1}\fi
\ifx \showISBNx    \undefined \def \showISBNx     #1{\unskip}     \fi
\ifx \showISBNxiii \undefined \def \showISBNxiii  #1{\unskip}     \fi
\ifx \showISSN     \undefined \def \showISSN      #1{\unskip}     \fi
\ifx \showLCCN     \undefined \def \showLCCN      #1{\unskip}     \fi
\ifx \shownote     \undefined \def \shownote      #1{#1}          \fi
\ifx \showarticletitle \undefined \def \showarticletitle #1{#1}   \fi
\ifx \showURL      \undefined \def \showURL       {\relax}        \fi
% invisible to TeX
\providecommand\bibfield[2]{#2}
\providecommand\bibinfo[2]{#2}
\providecommand\natexlab[1]{#1}
\providecommand\showeprint[2][]{arXiv:#2}

\bibitem[app(2023a)]%
        {appleaudio}
 \bibinfo{year}{2023}\natexlab{a}.
\newblock \bibinfo{title}{AirPods redefine the personal audio experience}.
\newblock
\newblock
\urldef\tempurl%
\url{https://www.apple.com/newsroom/2023/06/airpods-redefine-the-personal-audio-experience/}
\showURL{%
\tempurl}


\bibitem[app(2023b)]%
        {applevis}
 \bibinfo{year}{2023}\natexlab{b}.
\newblock \bibinfo{title}{AppleVis}.
\newblock
\newblock
\urldef\tempurl%
\url{https://www.applevis.com/}
\showURL{%
\tempurl}


\bibitem[bli(2023a)]%
        {blindreddit}
 \bibinfo{year}{2023}\natexlab{a}.
\newblock \bibinfo{title}{Blind and Visually Impaired Community}.
\newblock
\newblock
\urldef\tempurl%
\url{https://www.reddit.com/r/Blind/}
\showURL{%
\tempurl}


\bibitem[bli(2023b)]%
        {blindsquare}
 \bibinfo{year}{2023}\natexlab{b}.
\newblock \bibinfo{title}{BlindSquare}.
\newblock
\newblock
\urldef\tempurl%
\url{https://www.blindsquare.com/}
\showURL{%
\tempurl}


\bibitem[sou(2023)]%
        {soundscape}
 \bibinfo{year}{2023}\natexlab{}.
\newblock \bibinfo{title}{Microsoft Soundscape}.
\newblock
\newblock
\urldef\tempurl%
\url{https://www.microsoft.com/en-us/research/product/soundscape/}
\showURL{%
\tempurl}


\bibitem[Alexander~Wang(2024)]%
        {maringba}
\bibfield{author}{\bibinfo{person}{David~Lindlbauer. Alexander~Wang, Yi Fei~Cheng}.} \bibinfo{year}{2024}\natexlab{}.
\newblock \showarticletitle{MARingBA: Music-Adaptive Ringtones for Blended Audio Notification Delivery}. In \bibinfo{booktitle}{\emph{In Proceedings of the CHI Conference on Human Factors in Computing Systems (CHI '24)}}. \bibinfo{address}{Honolulu, HI,}.
\newblock
\urldef\tempurl%
\url{https://doi.org/10.1145/3613904.3642376}
\showURL{%
\tempurl}


\bibitem[Andr{\'e} et~al\mbox{.}(2012)]%
        {andre2012sound}
\bibfield{author}{\bibinfo{person}{C{\'e}dric~R Andr{\'e}}, \bibinfo{person}{Jean-Jacques Embrechts}, \bibinfo{person}{Jacques~G Verly}, \bibinfo{person}{Marc R{\'e}billat}, {and} \bibinfo{person}{Brian~FG Katz}.} \bibinfo{year}{2012}\natexlab{}.
\newblock \showarticletitle{Sound for 3D cinema and the sense of presence}. Georgia Institute of Technology.
\newblock


\bibitem[Ballou(1987)]%
        {ballou1987handbook}
\bibfield{author}{\bibinfo{person}{Greg Ballou}.} \bibinfo{year}{1987}\natexlab{}.
\newblock \showarticletitle{Handbook for Sound Engineers: The New Audio Cyclopedia, Howard W}.
\newblock \bibinfo{journal}{\emph{Sams and Co., Indianapolis}} (\bibinfo{year}{1987}).
\newblock


\bibitem[Bederson(1995)]%
        {bederson1995audio}
\bibfield{author}{\bibinfo{person}{Benjamin~B Bederson}.} \bibinfo{year}{1995}\natexlab{}.
\newblock \showarticletitle{Audio augmented reality: a prototype automated tour guide}. In \bibinfo{booktitle}{\emph{Conference companion on Human factors in computing systems}}. \bibinfo{pages}{210--211}.
\newblock


\bibitem[Blattner et~al\mbox{.}(1989)]%
        {blattner1989earcons}
\bibfield{author}{\bibinfo{person}{Meera~M Blattner}, \bibinfo{person}{Denise~A Sumikawa}, {and} \bibinfo{person}{Robert~M Greenberg}.} \bibinfo{year}{1989}\natexlab{}.
\newblock \showarticletitle{Earcons and icons: Their structure and common design principles}.
\newblock \bibinfo{journal}{\emph{Human--Computer Interaction}} \bibinfo{volume}{4}, \bibinfo{number}{1} (\bibinfo{year}{1989}), \bibinfo{pages}{11--44}.
\newblock


\bibitem[Brinkman et~al\mbox{.}(2015)]%
        {brinkman2015effect}
\bibfield{author}{\bibinfo{person}{Willem-Paul Brinkman}, \bibinfo{person}{Allart~RD Hoekstra}, {and} \bibinfo{person}{Ren{\'e} van EGMOND}.} \bibinfo{year}{2015}\natexlab{}.
\newblock \showarticletitle{The effect of 3D audio and other audio techniques on virtual reality experience}.
\newblock \bibinfo{journal}{\emph{Annual Review of Cybertherapy and Telemedicine 2015}} (\bibinfo{year}{2015}), \bibinfo{pages}{44--48}.
\newblock


\bibitem[Chang et~al\mbox{.}(2022)]%
        {omniscribe}
\bibfield{author}{\bibinfo{person}{Ruei-Che Chang}, \bibinfo{person}{Chao-Hsien Ting}, \bibinfo{person}{Chia-Sheng Hung}, \bibinfo{person}{Wan-Chen Lee}, \bibinfo{person}{Liang-Jin Chen}, \bibinfo{person}{Yu-Tzu Chao}, \bibinfo{person}{Bing-Yu Chen}, {and} \bibinfo{person}{Anhong Guo}.} \bibinfo{year}{2022}\natexlab{}.
\newblock \showarticletitle{OmniScribe: Authoring Immersive Audio Descriptions for 360° Videos}. In \bibinfo{booktitle}{\emph{Proceedings of the 35th Annual ACM Symposium on User Interface Software and Technology}} (Bend, OR, USA) \emph{(\bibinfo{series}{UIST '22})}. \bibinfo{publisher}{Association for Computing Machinery}, \bibinfo{address}{New York, NY, USA}, Article \bibinfo{articleno}{15}, \bibinfo{numpages}{14}~pages.
\newblock
\showISBNx{9781450393201}
\urldef\tempurl%
\url{https://doi.org/10.1145/3526113.3545613}
\showDOI{\tempurl}


\bibitem[Chatterjee et~al\mbox{.}(2022)]%
        {clearbuds}
\bibfield{author}{\bibinfo{person}{Ishan Chatterjee}, \bibinfo{person}{Maruchi Kim}, \bibinfo{person}{Vivek Jayaram}, \bibinfo{person}{Shyamnath Gollakota}, \bibinfo{person}{Ira Kemelmacher}, \bibinfo{person}{Shwetak Patel}, {and} \bibinfo{person}{Steven~M. Seitz}.} \bibinfo{year}{2022}\natexlab{}.
\newblock \showarticletitle{ClearBuds: wireless binaural earbuds for learning-based speech enhancement}. In \bibinfo{booktitle}{\emph{Proceedings of the 20th Annual International Conference on Mobile Systems, Applications and Services}} (Portland, Oregon) \emph{(\bibinfo{series}{MobiSys '22})}. \bibinfo{publisher}{Association for Computing Machinery}, \bibinfo{address}{New York, NY, USA}, \bibinfo{pages}{384–396}.
\newblock
\showISBNx{9781450391856}
\urldef\tempurl%
\url{https://doi.org/10.1145/3498361.3538933}
\showDOI{\tempurl}


\bibitem[Chen et~al\mbox{.}(2022)]%
        {opportune}
\bibfield{author}{\bibinfo{person}{Kuan-Wen Chen}, \bibinfo{person}{Yung-Ju Chang}, {and} \bibinfo{person}{Liwei Chan}.} \bibinfo{year}{2022}\natexlab{}.
\newblock \showarticletitle{Predicting Opportune Moments to Deliver Notifications in Virtual Reality}. In \bibinfo{booktitle}{\emph{Proceedings of the 2022 CHI Conference on Human Factors in Computing Systems}} (New Orleans, LA, USA) \emph{(\bibinfo{series}{CHI '22})}. \bibinfo{publisher}{Association for Computing Machinery}, \bibinfo{address}{New York, NY, USA}, Article \bibinfo{articleno}{186}, \bibinfo{numpages}{18}~pages.
\newblock
\showISBNx{9781450391573}
\urldef\tempurl%
\url{https://doi.org/10.1145/3491102.3517529}
\showDOI{\tempurl}


\bibitem[Clarke et~al\mbox{.}(2015)]%
        {clarke2015thematic}
\bibfield{author}{\bibinfo{person}{Victoria Clarke}, \bibinfo{person}{Virginia Braun}, {and} \bibinfo{person}{Nikki Hayfield}.} \bibinfo{year}{2015}\natexlab{}.
\newblock \showarticletitle{Thematic analysis}.
\newblock \bibinfo{journal}{\emph{Qualitative psychology: A practical guide to research methods}}  \bibinfo{volume}{3} (\bibinfo{year}{2015}), \bibinfo{pages}{222--248}.
\newblock


\bibitem[Cliffe et~al\mbox{.}(2021)]%
        {cliffe2021materialising}
\bibfield{author}{\bibinfo{person}{Laurence Cliffe}, \bibinfo{person}{James Mansell}, \bibinfo{person}{Chris Greenhalgh}, {and} \bibinfo{person}{Adrian Hazzard}.} \bibinfo{year}{2021}\natexlab{}.
\newblock \showarticletitle{Materialising contexts: virtual soundscapes for real-world exploration}.
\newblock \bibinfo{journal}{\emph{Personal and Ubiquitous Computing}}  \bibinfo{volume}{25} (\bibinfo{year}{2021}), \bibinfo{pages}{623--636}.
\newblock


\bibitem[Daiber et~al\mbox{.}(2021)]%
        {everydayobject}
\bibfield{author}{\bibinfo{person}{Florian Daiber}, \bibinfo{person}{Donald Degraen}, \bibinfo{person}{Andr\'{e} Zenner}, \bibinfo{person}{Tanja D\"{o}ring}, \bibinfo{person}{Frank Steinicke}, \bibinfo{person}{Oscar~Javier Ariza~Nunez}, {and} \bibinfo{person}{Adalberto~L. Simeone}.} \bibinfo{year}{2021}\natexlab{}.
\newblock \showarticletitle{Everyday Proxy Objects for Virtual Reality}. In \bibinfo{booktitle}{\emph{Extended Abstracts of the 2021 CHI Conference on Human Factors in Computing Systems}} (Yokohama, Japan) \emph{(\bibinfo{series}{CHI EA '21})}. \bibinfo{publisher}{Association for Computing Machinery}, \bibinfo{address}{New York, NY, USA}, Article \bibinfo{articleno}{101}, \bibinfo{numpages}{6}~pages.
\newblock
\showISBNx{9781450380959}
\urldef\tempurl%
\url{https://doi.org/10.1145/3411763.3441343}
\showDOI{\tempurl}


\bibitem[D'Atri et~al\mbox{.}(2007)]%
        {d2007system}
\bibfield{author}{\bibinfo{person}{Edoardo D'Atri}, \bibinfo{person}{Carlo~Maria Medaglia}, \bibinfo{person}{Alexandru Serbanati}, \bibinfo{person}{Ugo~Biader Ceipidor}, \bibinfo{person}{Emanuele Panizzi}, {and} \bibinfo{person}{Alessandro D'Atri}.} \bibinfo{year}{2007}\natexlab{}.
\newblock \showarticletitle{A system to aid blind people in the mobility: A usability test and its results}. In \bibinfo{booktitle}{\emph{Second International Conference on Systems (ICONS'07)}}. IEEE, \bibinfo{publisher}{Institute of Electrical and Electronics Engineers}, \bibinfo{address}{New York, NY, USA}, \bibinfo{pages}{35--35}.
\newblock


\bibitem[Endo et~al\mbox{.}(2021)]%
        {modularhmd}
\bibfield{author}{\bibinfo{person}{Isamu Endo}, \bibinfo{person}{Kazuki Takashima}, \bibinfo{person}{Maakito Inoue}, \bibinfo{person}{Kazuyuki Fujita}, \bibinfo{person}{Kiyoshi Kiyokawa}, {and} \bibinfo{person}{Yoshifumi Kitamura}.} \bibinfo{year}{2021}\natexlab{}.
\newblock \showarticletitle{ModularHMD: A Reconfigurable Mobile Head-Mounted Display Enabling Ad-Hoc Peripheral Interactions with the Real World}. In \bibinfo{booktitle}{\emph{The 34th Annual ACM Symposium on User Interface Software and Technology}} (Virtual Event, USA) \emph{(\bibinfo{series}{UIST '21})}. \bibinfo{publisher}{Association for Computing Machinery}, \bibinfo{address}{New York, NY, USA}, \bibinfo{pages}{100–117}.
\newblock
\showISBNx{9781450386357}
\urldef\tempurl%
\url{https://doi.org/10.1145/3472749.3474738}
\showDOI{\tempurl}


\bibitem[for Disease~Control and Prevention(2023)]%
        {normalnoise}
\bibfield{author}{\bibinfo{person}{Centers for Disease~Control} {and} \bibinfo{person}{Prevention}.} \bibinfo{year}{2023}\natexlab{}.
\newblock \bibinfo{title}{What Noises Cause Hearing Loss?}
\newblock
\newblock
\urldef\tempurl%
\url{https://www.cdc.gov/nceh/hearing_loss/what_noises_cause_hearing_loss.html}
\showURL{%
\tempurl}


\bibitem[Gaver and Norman(1988)]%
        {gaver1988everyday}
\bibfield{author}{\bibinfo{person}{William~W Gaver} {and} \bibinfo{person}{Donald~A Norman}.} \bibinfo{year}{1988}\natexlab{}.
\newblock \emph{\bibinfo{title}{Everyday listening and auditory icons}}.
\newblock \bibinfo{thesistype}{Ph.\,D. Dissertation}. \bibinfo{school}{University of California, San Diego, Department of Cognitive Science and Psychology}.
\newblock


\bibitem[Ghosh et~al\mbox{.}(2018)]%
        {notifivr}
\bibfield{author}{\bibinfo{person}{Sarthak Ghosh}, \bibinfo{person}{Lauren Winston}, \bibinfo{person}{Nishant Panchal}, \bibinfo{person}{Philippe Kimura-Thollander}, \bibinfo{person}{Jeff Hotnog}, \bibinfo{person}{Douglas Cheong}, \bibinfo{person}{Gabriel Reyes}, {and} \bibinfo{person}{Gregory~D. Abowd}.} \bibinfo{year}{2018}\natexlab{}.
\newblock \showarticletitle{NotifiVR: Exploring Interruptions and Notifications in Virtual Reality}.
\newblock \bibinfo{journal}{\emph{IEEE Transactions on Visualization and Computer Graphics}} \bibinfo{volume}{24}, \bibinfo{number}{4} (\bibinfo{year}{2018}), \bibinfo{pages}{1447--1456}.
\newblock
\urldef\tempurl%
\url{https://doi.org/10.1109/TVCG.2018.2793698}
\showDOI{\tempurl}


\bibitem[Gilkey and Weisenberger(1995)]%
        {gilkey1995sense}
\bibfield{author}{\bibinfo{person}{Robert~H Gilkey} {and} \bibinfo{person}{Janet~M Weisenberger}.} \bibinfo{year}{1995}\natexlab{}.
\newblock \showarticletitle{The sense of presence for the suddenly deafened adult: Implications for virtual environments}.
\newblock \bibinfo{journal}{\emph{Presence: Teleoperators \& Virtual Environments}} \bibinfo{volume}{4}, \bibinfo{number}{4} (\bibinfo{year}{1995}), \bibinfo{pages}{357--363}.
\newblock


\bibitem[Gruenefeld et~al\mbox{.}(2022)]%
        {vrception}
\bibfield{author}{\bibinfo{person}{Uwe Gruenefeld}, \bibinfo{person}{Jonas Auda}, \bibinfo{person}{Florian Mathis}, \bibinfo{person}{Stefan Schneegass}, \bibinfo{person}{Mohamed Khamis}, \bibinfo{person}{Jan Gugenheimer}, {and} \bibinfo{person}{Sven Mayer}.} \bibinfo{year}{2022}\natexlab{}.
\newblock \showarticletitle{VRception: Rapid Prototyping of Cross-Reality Systems in Virtual Reality}. In \bibinfo{booktitle}{\emph{Proceedings of the 2022 CHI Conference on Human Factors in Computing Systems}} (New Orleans, LA, USA) \emph{(\bibinfo{series}{CHI '22})}. \bibinfo{publisher}{Association for Computing Machinery}, \bibinfo{address}{New York, NY, USA}, Article \bibinfo{articleno}{611}, \bibinfo{numpages}{15}~pages.
\newblock
\showISBNx{9781450391573}
\urldef\tempurl%
\url{https://doi.org/10.1145/3491102.3501821}
\showDOI{\tempurl}


\bibitem[Haas et~al\mbox{.}(2018)]%
        {curation}
\bibfield{author}{\bibinfo{person}{Gabriel Haas}, \bibinfo{person}{Evgeny Stemasov}, {and} \bibinfo{person}{Enrico Rukzio}.} \bibinfo{year}{2018}\natexlab{}.
\newblock \showarticletitle{Can't You Hear Me? Investigating Personal Soundscape Curation}. In \bibinfo{booktitle}{\emph{Proceedings of the 17th International Conference on Mobile and Ubiquitous Multimedia}} (Cairo, Egypt) \emph{(\bibinfo{series}{MUM '18})}. \bibinfo{publisher}{Association for Computing Machinery}, \bibinfo{address}{New York, NY, USA}, \bibinfo{pages}{59–69}.
\newblock
\showISBNx{9781450365949}
\urldef\tempurl%
\url{https://doi.org/10.1145/3282894.3282897}
\showDOI{\tempurl}


\bibitem[Hagood(2011)]%
        {hagood2011quiet}
\bibfield{author}{\bibinfo{person}{Mack Hagood}.} \bibinfo{year}{2011}\natexlab{}.
\newblock \showarticletitle{Quiet comfort: Noise, otherness, and the mobile production of personal space}.
\newblock \bibinfo{journal}{\emph{American Quarterly}} \bibinfo{volume}{63}, \bibinfo{number}{3} (\bibinfo{year}{2011}), \bibinfo{pages}{573--589}.
\newblock


\bibitem[Hart(2006)]%
        {nasatlx}
\bibfield{author}{\bibinfo{person}{Sandra~G Hart}.} \bibinfo{year}{2006}\natexlab{}.
\newblock \showarticletitle{NASA-task load index (NASA-TLX); 20 years later}. In \bibinfo{booktitle}{\emph{Proceedings of the human factors and ergonomics society annual meeting}}, Vol.~\bibinfo{volume}{50}. Sage publications Sage CA: Los Angeles, CA, \bibinfo{pages}{904--908}.
\newblock


\bibitem[Hartmann et~al\mbox{.}(2019)]%
        {realitycheck}
\bibfield{author}{\bibinfo{person}{Jeremy Hartmann}, \bibinfo{person}{Christian Holz}, \bibinfo{person}{Eyal Ofek}, {and} \bibinfo{person}{Andrew~D. Wilson}.} \bibinfo{year}{2019}\natexlab{}.
\newblock \showarticletitle{RealityCheck: Blending Virtual Environments with Situated Physical Reality}. In \bibinfo{booktitle}{\emph{Proceedings of the 2019 CHI Conference on Human Factors in Computing Systems}} (Glasgow, Scotland Uk) \emph{(\bibinfo{series}{CHI '19})}. \bibinfo{publisher}{Association for Computing Machinery}, \bibinfo{address}{New York, NY, USA}, \bibinfo{pages}{1–12}.
\newblock
\showISBNx{9781450359702}
\urldef\tempurl%
\url{https://doi.org/10.1145/3290605.3300577}
\showDOI{\tempurl}


\bibitem[Herskovitz et~al\mbox{.}(2020)]%
        {mobilear}
\bibfield{author}{\bibinfo{person}{Jaylin Herskovitz}, \bibinfo{person}{Jason Wu}, \bibinfo{person}{Samuel White}, \bibinfo{person}{Amy Pavel}, \bibinfo{person}{Gabriel Reyes}, \bibinfo{person}{Anhong Guo}, {and} \bibinfo{person}{Jeffrey~P. Bigham}.} \bibinfo{year}{2020}\natexlab{}.
\newblock \showarticletitle{Making Mobile Augmented Reality Applications Accessible}. In \bibinfo{booktitle}{\emph{The 22nd International ACM SIGACCESS Conference on Computers and Accessibility}} (Virtual Event, Greece) \emph{(\bibinfo{series}{ASSETS '20})}. \bibinfo{publisher}{Association for Computing Machinery}, \bibinfo{address}{New York, NY, USA}, Article \bibinfo{articleno}{3}, \bibinfo{numpages}{14}~pages.
\newblock
\showISBNx{9781450371032}
\urldef\tempurl%
\url{https://doi.org/10.1145/3373625.3417006}
\showDOI{\tempurl}


\bibitem[Hettiarachchi and Wigdor(2016)]%
        {annexingreality}
\bibfield{author}{\bibinfo{person}{Anuruddha Hettiarachchi} {and} \bibinfo{person}{Daniel Wigdor}.} \bibinfo{year}{2016}\natexlab{}.
\newblock \showarticletitle{Annexing Reality: Enabling Opportunistic Use of Everyday Objects as Tangible Proxies in Augmented Reality}. In \bibinfo{booktitle}{\emph{Proceedings of the 2016 CHI Conference on Human Factors in Computing Systems}} (San Jose, California, USA) \emph{(\bibinfo{series}{CHI '16})}. \bibinfo{publisher}{Association for Computing Machinery}, \bibinfo{address}{New York, NY, USA}, \bibinfo{pages}{1957–1967}.
\newblock
\showISBNx{9781450333627}
\urldef\tempurl%
\url{https://doi.org/10.1145/2858036.2858134}
\showDOI{\tempurl}


\bibitem[Hsieh et~al\mbox{.}(2020)]%
        {notiposition}
\bibfield{author}{\bibinfo{person}{Ching-Yu Hsieh}, \bibinfo{person}{Yi-Shyuan Chiang}, \bibinfo{person}{Hung-Yu Chiu}, {and} \bibinfo{person}{Yung-Ju Chang}.} \bibinfo{year}{2020}\natexlab{}.
\newblock \showarticletitle{Bridging the Virtual and Real Worlds: A Preliminary Study of Messaging Notifications in Virtual Reality}. In \bibinfo{booktitle}{\emph{Proceedings of the 2020 CHI Conference on Human Factors in Computing Systems}} (Honolulu, HI, USA) \emph{(\bibinfo{series}{CHI '20})}. \bibinfo{publisher}{Association for Computing Machinery}, \bibinfo{address}{New York, NY, USA}, \bibinfo{pages}{1–14}.
\newblock
\showISBNx{9781450367080}
\urldef\tempurl%
\url{https://doi.org/10.1145/3313831.3376228}
\showDOI{\tempurl}


\bibitem[Jain et~al\mbox{.}(2022)]%
        {protosound}
\bibfield{author}{\bibinfo{person}{Dhruv Jain}, \bibinfo{person}{Khoa Huynh Anh~Nguyen}, \bibinfo{person}{Steven M.~Goodman}, \bibinfo{person}{Rachel Grossman-Kahn}, \bibinfo{person}{Hung Ngo}, \bibinfo{person}{Aditya Kusupati}, \bibinfo{person}{Ruofei Du}, \bibinfo{person}{Alex Olwal}, \bibinfo{person}{Leah Findlater}, {and} \bibinfo{person}{Jon E.~Froehlich}.} \bibinfo{year}{2022}\natexlab{}.
\newblock \showarticletitle{ProtoSound: A Personalized and Scalable Sound Recognition System for Deaf and Hard-of-Hearing Users}. In \bibinfo{booktitle}{\emph{Proceedings of the 2022 CHI Conference on Human Factors in Computing Systems}} (New Orleans, LA, USA) \emph{(\bibinfo{series}{CHI '22})}. \bibinfo{publisher}{Association for Computing Machinery}, \bibinfo{address}{New York, NY, USA}, Article \bibinfo{articleno}{305}, \bibinfo{numpages}{16}~pages.
\newblock
\showISBNx{9781450391573}
\urldef\tempurl%
\url{https://doi.org/10.1145/3491102.3502020}
\showDOI{\tempurl}


\bibitem[Jain et~al\mbox{.}(2021)]%
        {soundvra11y}
\bibfield{author}{\bibinfo{person}{Dhruv Jain}, \bibinfo{person}{Sasa Junuzovic}, \bibinfo{person}{Eyal Ofek}, \bibinfo{person}{Mike Sinclair}, \bibinfo{person}{John R.~Porter}, \bibinfo{person}{Chris Yoon}, \bibinfo{person}{Swetha Machanavajhala}, {and} \bibinfo{person}{Meredith Ringel~Morris}.} \bibinfo{year}{2021}\natexlab{}.
\newblock \showarticletitle{Towards Sound Accessibility in Virtual Reality}. In \bibinfo{booktitle}{\emph{Proceedings of the 2021 International Conference on Multimodal Interaction}} (Montr\'{e}al, QC, Canada) \emph{(\bibinfo{series}{ICMI '21})}. \bibinfo{publisher}{Association for Computing Machinery}, \bibinfo{address}{New York, NY, USA}, \bibinfo{pages}{80–91}.
\newblock
\showISBNx{9781450384810}
\urldef\tempurl%
\url{https://doi.org/10.1145/3462244.3479946}
\showDOI{\tempurl}


\bibitem[Jain et~al\mbox{.}(2023)]%
        {acousticmapCSCW23}
\bibfield{author}{\bibinfo{person}{Gaurav Jain}, \bibinfo{person}{Yuanyang Teng}, \bibinfo{person}{Dong~Heon Cho}, \bibinfo{person}{Yunhao Xing}, \bibinfo{person}{Maryam Aziz}, {and} \bibinfo{person}{Brian~A. Smith}.} \bibinfo{year}{2023}\natexlab{}.
\newblock \showarticletitle{"I Want to Figure Things Out": Supporting Exploration in Navigation for People with Visual Impairments}.
\newblock \bibinfo{journal}{\emph{Proc. ACM Hum.-Comput. Interact.}} \bibinfo{volume}{7}, \bibinfo{number}{CSCW1}, Article \bibinfo{articleno}{63} (\bibinfo{date}{apr} \bibinfo{year}{2023}), \bibinfo{numpages}{28}~pages.
\newblock
\urldef\tempurl%
\url{https://doi.org/10.1145/3579496}
\showDOI{\tempurl}


\bibitem[Ji et~al\mbox{.}(2022)]%
        {vrbubble}
\bibfield{author}{\bibinfo{person}{Tiger~F. Ji}, \bibinfo{person}{Brianna Cochran}, {and} \bibinfo{person}{Yuhang Zhao}.} \bibinfo{year}{2022}\natexlab{}.
\newblock \showarticletitle{VRBubble: Enhancing Peripheral Awareness of Avatars for People with Visual Impairments in Social Virtual Reality}. In \bibinfo{booktitle}{\emph{Proceedings of the 24th International ACM SIGACCESS Conference on Computers and Accessibility}} (Athens, Greece) \emph{(\bibinfo{series}{ASSETS '22})}. \bibinfo{publisher}{Association for Computing Machinery}, \bibinfo{address}{New York, NY, USA}, Article \bibinfo{articleno}{3}, \bibinfo{numpages}{17}~pages.
\newblock
\showISBNx{9781450392587}
\urldef\tempurl%
\url{https://doi.org/10.1145/3517428.3544821}
\showDOI{\tempurl}


\bibitem[Johansen et~al\mbox{.}(2022)]%
        {soundscapesurvey}
\bibfield{author}{\bibinfo{person}{Stine~S. Johansen}, \bibinfo{person}{Niels van Berkel}, {and} \bibinfo{person}{Jonas Fritsch}.} \bibinfo{year}{2022}\natexlab{}.
\newblock \showarticletitle{Characterising Soundscape Research in Human-Computer Interaction}. In \bibinfo{booktitle}{\emph{Designing Interactive Systems Conference}} (Virtual Event, Australia) \emph{(\bibinfo{series}{DIS '22})}. \bibinfo{publisher}{Association for Computing Machinery}, \bibinfo{address}{New York, NY, USA}, \bibinfo{pages}{1394–1417}.
\newblock
\showISBNx{9781450393584}
\urldef\tempurl%
\url{https://doi.org/10.1145/3532106.3533458}
\showDOI{\tempurl}


\bibitem[Kari et~al\mbox{.}(2021)]%
        {soundsride}
\bibfield{author}{\bibinfo{person}{Mohamed Kari}, \bibinfo{person}{Tobias Grosse-Puppendahl}, \bibinfo{person}{Alexander Jagaciak}, \bibinfo{person}{David Bethge}, \bibinfo{person}{Reinhard Sch\"{u}tte}, {and} \bibinfo{person}{Christian Holz}.} \bibinfo{year}{2021}\natexlab{}.
\newblock \showarticletitle{SoundsRide: Affordance-Synchronized Music Mixing for In-Car Audio Augmented Reality}. In \bibinfo{booktitle}{\emph{The 34th Annual ACM Symposium on User Interface Software and Technology}} (Virtual Event, USA) \emph{(\bibinfo{series}{UIST '21})}. \bibinfo{publisher}{Association for Computing Machinery}, \bibinfo{address}{New York, NY, USA}, \bibinfo{pages}{118–133}.
\newblock
\showISBNx{9781450386357}
\urldef\tempurl%
\url{https://doi.org/10.1145/3472749.3474739}
\showDOI{\tempurl}


\bibitem[Klemmer et~al\mbox{.}(2000)]%
        {suede}
\bibfield{author}{\bibinfo{person}{Scott~R. Klemmer}, \bibinfo{person}{Anoop~K. Sinha}, \bibinfo{person}{Jack Chen}, \bibinfo{person}{James~A. Landay}, \bibinfo{person}{Nadeem Aboobaker}, {and} \bibinfo{person}{Annie Wang}.} \bibinfo{year}{2000}\natexlab{}.
\newblock \showarticletitle{Suede: A Wizard of Oz Prototyping Tool for Speech User Interfaces}. In \bibinfo{booktitle}{\emph{Proceedings of the 13th Annual ACM Symposium on User Interface Software and Technology}} (San Diego, California, USA) \emph{(\bibinfo{series}{UIST '00})}. \bibinfo{publisher}{Association for Computing Machinery}, \bibinfo{address}{New York, NY, USA}, \bibinfo{pages}{1–10}.
\newblock
\showISBNx{1581132123}
\urldef\tempurl%
\url{https://doi.org/10.1145/354401.354406}
\showDOI{\tempurl}


\bibitem[Kreiss et~al\mbox{.}(2022)]%
        {kreiss2022context}
\bibfield{author}{\bibinfo{person}{Elisa Kreiss}, \bibinfo{person}{Cynthia Bennett}, \bibinfo{person}{Shayan Hooshmand}, \bibinfo{person}{Eric Zelikman}, \bibinfo{person}{Meredith~Ringel Morris}, {and} \bibinfo{person}{Christopher Potts}.} \bibinfo{year}{2022}\natexlab{}.
\newblock \showarticletitle{Context Matters for Image Descriptions for Accessibility: Challenges for Referenceless Evaluation Metrics}.
\newblock \bibinfo{journal}{\emph{arXiv preprint arXiv:2205.10646}} (\bibinfo{year}{2022}).
\newblock


\bibitem[Krzyzaniak et~al\mbox{.}(2019)]%
        {sixtypes}
\bibfield{author}{\bibinfo{person}{Michael Krzyzaniak}, \bibinfo{person}{David Frohlich}, {and} \bibinfo{person}{Philip~J.B. Jackson}.} \bibinfo{year}{2019}\natexlab{}.
\newblock \showarticletitle{Six types of audio that DEFY reality! A taxonomy of audio augmented reality with examples}. In \bibinfo{booktitle}{\emph{Proceedings of the 14th International Audio Mostly Conference: A Journey in Sound}} (Nottingham, United Kingdom) \emph{(\bibinfo{series}{AM '19})}. \bibinfo{publisher}{Association for Computing Machinery}, \bibinfo{address}{New York, NY, USA}, \bibinfo{pages}{160–167}.
\newblock
\showISBNx{9781450372978}
\urldef\tempurl%
\url{https://doi.org/10.1145/3356590.3356615}
\showDOI{\tempurl}


\bibitem[L.~Franz et~al\mbox{.}(2021)]%
        {nearmi}
\bibfield{author}{\bibinfo{person}{Rachel L.~Franz}, \bibinfo{person}{Sasa Junuzovic}, {and} \bibinfo{person}{Martez Mott}.} \bibinfo{year}{2021}\natexlab{}.
\newblock \showarticletitle{Nearmi: A Framework for Designing Point of Interest Techniques for VR Users with Limited Mobility}. In \bibinfo{booktitle}{\emph{Proceedings of the 23rd International ACM SIGACCESS Conference on Computers and Accessibility}} (Virtual Event, USA) \emph{(\bibinfo{series}{ASSETS '21})}. \bibinfo{publisher}{Association for Computing Machinery}, \bibinfo{address}{New York, NY, USA}, Article \bibinfo{articleno}{5}, \bibinfo{numpages}{14}~pages.
\newblock
\showISBNx{9781450383066}
\urldef\tempurl%
\url{https://doi.org/10.1145/3441852.3471230}
\showDOI{\tempurl}


\bibitem[Laput et~al\mbox{.}(2018)]%
        {ubicoustics}
\bibfield{author}{\bibinfo{person}{Gierad Laput}, \bibinfo{person}{Karan Ahuja}, \bibinfo{person}{Mayank Goel}, {and} \bibinfo{person}{Chris Harrison}.} \bibinfo{year}{2018}\natexlab{}.
\newblock \showarticletitle{Ubicoustics: Plug-and-Play Acoustic Activity Recognition}. In \bibinfo{booktitle}{\emph{Proceedings of the 31st Annual ACM Symposium on User Interface Software and Technology}} (Berlin, Germany) \emph{(\bibinfo{series}{UIST '18})}. \bibinfo{publisher}{Association for Computing Machinery}, \bibinfo{address}{New York, NY, USA}, \bibinfo{pages}{213–224}.
\newblock
\showISBNx{9781450359481}
\urldef\tempurl%
\url{https://doi.org/10.1145/3242587.3242609}
\showDOI{\tempurl}


\bibitem[Larsson et~al\mbox{.}(2010)]%
        {larsson2010auditory}
\bibfield{author}{\bibinfo{person}{Pontus Larsson}, \bibinfo{person}{Aleksander V{\"a}ljam{\"a}e}, \bibinfo{person}{Daniel V{\"a}stfj{\"a}ll}, \bibinfo{person}{Ana Tajadura-Jim{\'e}nez}, {and} \bibinfo{person}{Mendel Kleiner}.} \bibinfo{year}{2010}\natexlab{}.
\newblock \showarticletitle{Auditory-induced presence in mixed reality environments and related technology}.
\newblock In \bibinfo{booktitle}{\emph{The engineering of mixed reality systems}}. \bibinfo{publisher}{Springer}, \bibinfo{pages}{143--163}.
\newblock


\bibitem[Lecuyer et~al\mbox{.}(2003)]%
        {lecuyer2003homere}
\bibfield{author}{\bibinfo{person}{A. Lecuyer}, \bibinfo{person}{P. Mobuchon}, \bibinfo{person}{C. Megard}, \bibinfo{person}{J. Perret}, \bibinfo{person}{C. Andriot}, {and} \bibinfo{person}{J.-P. Colinot}.} \bibinfo{year}{2003}\natexlab{}.
\newblock \showarticletitle{HOMERE: a multimodal system for visually impaired people to explore virtual environments}. In \bibinfo{booktitle}{\emph{IEEE Virtual Reality, 2003. Proceedings.}} \bibinfo{publisher}{Institute of Electrical and Electronics Engineers}, \bibinfo{address}{New York, NY, USA}, \bibinfo{pages}{251--258}.
\newblock
\urldef\tempurl%
\url{https://doi.org/10.1109/VR.2003.1191147}
\showDOI{\tempurl}


\bibitem[Lee et~al\mbox{.}(2022b)]%
        {collabally}
\bibfield{author}{\bibinfo{person}{Cheuk Yin~Phipson Lee}, \bibinfo{person}{Zhuohao Zhang}, \bibinfo{person}{Jaylin Herskovitz}, \bibinfo{person}{JooYoung Seo}, {and} \bibinfo{person}{Anhong Guo}.} \bibinfo{year}{2022}\natexlab{b}.
\newblock \showarticletitle{CollabAlly: Accessible Collaboration Awareness in Document Editing}. In \bibinfo{booktitle}{\emph{Proceedings of the 2022 CHI Conference on Human Factors in Computing Systems}} (New Orleans, LA, USA) \emph{(\bibinfo{series}{CHI '22})}. \bibinfo{publisher}{Association for Computing Machinery}, \bibinfo{address}{New York, NY, USA}, Article \bibinfo{articleno}{596}, \bibinfo{numpages}{17}~pages.
\newblock
\showISBNx{9781450391573}
\urldef\tempurl%
\url{https://doi.org/10.1145/3491102.3517635}
\showDOI{\tempurl}


\bibitem[Lee et~al\mbox{.}(2022a)]%
        {imageexplorer}
\bibfield{author}{\bibinfo{person}{Jaewook Lee}, \bibinfo{person}{Jaylin Herskovitz}, \bibinfo{person}{Yi-Hao Peng}, {and} \bibinfo{person}{Anhong Guo}.} \bibinfo{year}{2022}\natexlab{a}.
\newblock \showarticletitle{ImageExplorer: Multi-Layered Touch Exploration to Encourage Skepticism Towards Imperfect AI-Generated Image Captions}. In \bibinfo{booktitle}{\emph{Proceedings of the 2022 CHI Conference on Human Factors in Computing Systems}} (New Orleans, LA, USA) \emph{(\bibinfo{series}{CHI '22})}. \bibinfo{publisher}{Association for Computing Machinery}, \bibinfo{address}{New York, NY, USA}, Article \bibinfo{articleno}{462}, \bibinfo{numpages}{15}~pages.
\newblock
\showISBNx{9781450391573}
\urldef\tempurl%
\url{https://doi.org/10.1145/3491102.3501966}
\showDOI{\tempurl}


\bibitem[Lee et~al\mbox{.}(2009)]%
        {lee2009snackbot}
\bibfield{author}{\bibinfo{person}{Min~Kyung Lee}, \bibinfo{person}{Jodi Forlizzi}, \bibinfo{person}{Paul~E Rybski}, \bibinfo{person}{Frederick Crabbe}, \bibinfo{person}{Wayne Chung}, \bibinfo{person}{Josh Finkle}, \bibinfo{person}{Eric Glaser}, {and} \bibinfo{person}{Sara Kiesler}.} \bibinfo{year}{2009}\natexlab{}.
\newblock \showarticletitle{The snackbot: documenting the design of a robot for long-term human-robot interaction}. In \bibinfo{booktitle}{\emph{Proceedings of the 4th ACM/IEEE international conference on Human robot interaction}}. \bibinfo{pages}{7--14}.
\newblock


\bibitem[Li et~al\mbox{.}(2022)]%
        {soundvizvr}
\bibfield{author}{\bibinfo{person}{Ziming Li}, \bibinfo{person}{Shannon Connell}, \bibinfo{person}{Wendy Dannels}, {and} \bibinfo{person}{Roshan Peiris}.} \bibinfo{year}{2022}\natexlab{}.
\newblock \showarticletitle{SoundVizVR: Sound Indicators for Accessible Sounds in Virtual Reality for Deaf or Hard-of-Hearing Users}. In \bibinfo{booktitle}{\emph{Proceedings of the 24th International ACM SIGACCESS Conference on Computers and Accessibility}} (Athens, Greece) \emph{(\bibinfo{series}{ASSETS '22})}. \bibinfo{publisher}{Association for Computing Machinery}, \bibinfo{address}{New York, NY, USA}, Article \bibinfo{articleno}{5}, \bibinfo{numpages}{13}~pages.
\newblock
\showISBNx{9781450392587}
\urldef\tempurl%
\url{https://doi.org/10.1145/3517428.3544817}
\showDOI{\tempurl}


\bibitem[Lindlbauer and Wilson(2018)]%
        {remixedreality}
\bibfield{author}{\bibinfo{person}{David Lindlbauer} {and} \bibinfo{person}{Andy~D. Wilson}.} \bibinfo{year}{2018}\natexlab{}.
\newblock \showarticletitle{Remixed Reality: Manipulating Space and Time in Augmented Reality}. In \bibinfo{booktitle}{\emph{Proceedings of the 2018 CHI Conference on Human Factors in Computing Systems}} (Montreal QC, Canada) \emph{(\bibinfo{series}{CHI '18})}. \bibinfo{publisher}{Association for Computing Machinery}, \bibinfo{address}{New York, NY, USA}, \bibinfo{pages}{1–13}.
\newblock
\showISBNx{9781450356206}
\urldef\tempurl%
\url{https://doi.org/10.1145/3173574.3173703}
\showDOI{\tempurl}


\bibitem[Lindstrom and Bates(1988)]%
        {lindstrom1988newton}
\bibfield{author}{\bibinfo{person}{Mary~J Lindstrom} {and} \bibinfo{person}{Douglas~M Bates}.} \bibinfo{year}{1988}\natexlab{}.
\newblock \showarticletitle{Newton—Raphson and EM algorithms for linear mixed-effects models for repeated-measures data}.
\newblock \bibinfo{journal}{\emph{J. Amer. Statist. Assoc.}} \bibinfo{volume}{83}, \bibinfo{number}{404} (\bibinfo{year}{1988}), \bibinfo{pages}{1014--1022}.
\newblock


\bibitem[Lombard and Ditton(1997)]%
        {lombard1997heart}
\bibfield{author}{\bibinfo{person}{Matthew Lombard} {and} \bibinfo{person}{Theresa Ditton}.} \bibinfo{year}{1997}\natexlab{}.
\newblock \showarticletitle{At the heart of it all: The concept of presence}.
\newblock \bibinfo{journal}{\emph{Journal of computer-mediated communication}} \bibinfo{volume}{3}, \bibinfo{number}{2} (\bibinfo{year}{1997}), \bibinfo{pages}{JCMC321}.
\newblock


\bibitem[Lopez et~al\mbox{.}(2018)]%
        {lopez2018audio}
\bibfield{author}{\bibinfo{person}{Mariana Lopez}, \bibinfo{person}{Gavin Kearney}, {and} \bibinfo{person}{Kriszti{\'a}n Hofst{\"a}dter}.} \bibinfo{year}{2018}\natexlab{}.
\newblock \showarticletitle{Audio Description in the UK: What works, what doesn’t, and understanding the need for personalising access}.
\newblock \bibinfo{journal}{\emph{British journal of visual impairment}} \bibinfo{volume}{36}, \bibinfo{number}{3} (\bibinfo{year}{2018}), \bibinfo{pages}{274--291}.
\newblock


\bibitem[Lopez et~al\mbox{.}(2022)]%
        {lopez2022seeing}
\bibfield{author}{\bibinfo{person}{Mariana Lopez}, \bibinfo{person}{Gavin Kearney}, {and} \bibinfo{person}{Kriszti{\'a}n Hofst{\"a}dter}.} \bibinfo{year}{2022}\natexlab{}.
\newblock \showarticletitle{Seeing films through sound: Sound design, spatial audio, and accessibility for visually impaired audiences}.
\newblock \bibinfo{journal}{\emph{British Journal of Visual Impairment}} \bibinfo{volume}{40}, \bibinfo{number}{2} (\bibinfo{year}{2022}), \bibinfo{pages}{117--144}.
\newblock


\bibitem[Mamuji et~al\mbox{.}(2005)]%
        {mamuji2005attentive}
\bibfield{author}{\bibinfo{person}{Aadil Mamuji}, \bibinfo{person}{Roel Vertegaal}, \bibinfo{person}{Changuk Sohn}, {and} \bibinfo{person}{Daniel Cheng}.} \bibinfo{year}{2005}\natexlab{}.
\newblock \showarticletitle{Attentive Headphones: Augmenting Conversational Attention with a Real World TiVo}. In \bibinfo{booktitle}{\emph{Extended Abstracts of CHI}}, Vol.~\bibinfo{volume}{5}.
\newblock


\bibitem[McGill et~al\mbox{.}(2015)]%
        {doseofreality}
\bibfield{author}{\bibinfo{person}{Mark McGill}, \bibinfo{person}{Daniel Boland}, \bibinfo{person}{Roderick Murray-Smith}, {and} \bibinfo{person}{Stephen Brewster}.} \bibinfo{year}{2015}\natexlab{}.
\newblock \showarticletitle{A Dose of Reality: Overcoming Usability Challenges in VR Head-Mounted Displays}. In \bibinfo{booktitle}{\emph{Proceedings of the 33rd Annual ACM Conference on Human Factors in Computing Systems}} (Seoul, Republic of Korea) \emph{(\bibinfo{series}{CHI '15})}. \bibinfo{publisher}{Association for Computing Machinery}, \bibinfo{address}{New York, NY, USA}, \bibinfo{pages}{2143–2152}.
\newblock
\showISBNx{9781450331456}
\urldef\tempurl%
\url{https://doi.org/10.1145/2702123.2702382}
\showDOI{\tempurl}


\bibitem[McGill et~al\mbox{.}(2020)]%
        {mcgill}
\bibfield{author}{\bibinfo{person}{Mark McGill}, \bibinfo{person}{Stephen Brewster}, \bibinfo{person}{David McGookin}, {and} \bibinfo{person}{Graham Wilson}.} \bibinfo{year}{2020}\natexlab{}.
\newblock \showarticletitle{Acoustic Transparency and the Changing Soundscape of Auditory Mixed Reality}. In \bibinfo{booktitle}{\emph{Proceedings of the 2020 CHI Conference on Human Factors in Computing Systems}} (Honolulu, HI, USA) \emph{(\bibinfo{series}{CHI '20})}. \bibinfo{publisher}{Association for Computing Machinery}, \bibinfo{address}{New York, NY, USA}, \bibinfo{pages}{1–16}.
\newblock
\showISBNx{9781450367080}
\urldef\tempurl%
\url{https://doi.org/10.1145/3313831.3376702}
\showDOI{\tempurl}


\bibitem[Milgram and Kishino(1994)]%
        {milgram1994taxonomy}
\bibfield{author}{\bibinfo{person}{Paul Milgram} {and} \bibinfo{person}{Fumio Kishino}.} \bibinfo{year}{1994}\natexlab{}.
\newblock \showarticletitle{A taxonomy of mixed reality visual displays}.
\newblock \bibinfo{journal}{\emph{IEICE TRANSACTIONS on Information and Systems}} \bibinfo{volume}{77}, \bibinfo{number}{12} (\bibinfo{year}{1994}), \bibinfo{pages}{1321--1329}.
\newblock


\bibitem[Mott et~al\mbox{.}(2019)]%
        {a11ybydesign}
\bibfield{author}{\bibinfo{person}{Martez Mott}, \bibinfo{person}{Edward Cutrell}, \bibinfo{person}{Mar Gonzalez~Franco}, \bibinfo{person}{Christian Holz}, \bibinfo{person}{Eyal Ofek}, \bibinfo{person}{Richard Stoakley}, {and} \bibinfo{person}{Meredith Ringel~Morris}.} \bibinfo{year}{2019}\natexlab{}.
\newblock \showarticletitle{Accessible by Design: An Opportunity for Virtual Reality}. In \bibinfo{booktitle}{\emph{2019 IEEE International Symposium on Mixed and Augmented Reality Adjunct (ISMAR-Adjunct)}}. \bibinfo{pages}{451--454}.
\newblock
\urldef\tempurl%
\url{https://doi.org/10.1109/ISMAR-Adjunct.2019.00122}
\showDOI{\tempurl}


\bibitem[Mott et~al\mbox{.}(2020)]%
        {limitedmobility}
\bibfield{author}{\bibinfo{person}{Martez Mott}, \bibinfo{person}{John Tang}, \bibinfo{person}{Shaun Kane}, \bibinfo{person}{Edward Cutrell}, {and} \bibinfo{person}{Meredith Ringel~Morris}.} \bibinfo{year}{2020}\natexlab{}.
\newblock \showarticletitle{“I Just Went into It Assuming That I Wouldn't Be Able to Have the Full Experience”: Understanding the Accessibility of Virtual Reality for People with Limited Mobility}. In \bibinfo{booktitle}{\emph{Proceedings of the 22nd International ACM SIGACCESS Conference on Computers and Accessibility}} (Virtual Event, Greece) \emph{(\bibinfo{series}{ASSETS '20})}. \bibinfo{publisher}{Association for Computing Machinery}, \bibinfo{address}{New York, NY, USA}, Article \bibinfo{articleno}{43}, \bibinfo{numpages}{13}~pages.
\newblock
\showISBNx{9781450371032}
\urldef\tempurl%
\url{https://doi.org/10.1145/3373625.3416998}
\showDOI{\tempurl}


\bibitem[Murray et~al\mbox{.}(2000)]%
        {murray2000presence}
\bibfield{author}{\bibinfo{person}{Craig~D Murray}, \bibinfo{person}{Paul Arnold}, {and} \bibinfo{person}{Ben Thornton}.} \bibinfo{year}{2000}\natexlab{}.
\newblock \showarticletitle{Presence accompanying induced hearing loss: Implications for immersive virtual environments}.
\newblock \bibinfo{journal}{\emph{Presence}} \bibinfo{volume}{9}, \bibinfo{number}{2} (\bibinfo{year}{2000}), \bibinfo{pages}{137--148}.
\newblock


\bibitem[Nair et~al\mbox{.}(2021a)]%
        {navstick}
\bibfield{author}{\bibinfo{person}{Vishnu Nair}, \bibinfo{person}{Jay~L Karp}, \bibinfo{person}{Samuel Silverman}, \bibinfo{person}{Mohar Kalra}, \bibinfo{person}{Hollis Lehv}, \bibinfo{person}{Faizan Jamil}, {and} \bibinfo{person}{Brian~A. Smith}.} \bibinfo{year}{2021}\natexlab{a}.
\newblock \showarticletitle{NavStick: Making Video Games Blind-Accessible via the Ability to Look Around}. In \bibinfo{booktitle}{\emph{The 34th Annual ACM Symposium on User Interface Software and Technology}}. \bibinfo{publisher}{Association for Computing Machinery}, \bibinfo{address}{New York, NY, USA}, \bibinfo{pages}{538–551}.
\newblock
\showISBNx{9781450386357}
\urldef\tempurl%
\url{https://doi.org/10.1145/3472749.3474768}
\showURL{%
\tempurl}


\bibitem[Nair et~al\mbox{.}(2022)]%
        {nair2022uncovering}
\bibfield{author}{\bibinfo{person}{Vishnu Nair}, \bibinfo{person}{Shao-en Ma}, \bibinfo{person}{Ricardo~E Gonzalez~Penuela}, \bibinfo{person}{Yicheng He}, \bibinfo{person}{Karen Lin}, \bibinfo{person}{Mason Hayes}, \bibinfo{person}{Hannah Huddleston}, \bibinfo{person}{Matthew Donnelly}, {and} \bibinfo{person}{Brian~A Smith}.} \bibinfo{year}{2022}\natexlab{}.
\newblock \showarticletitle{Uncovering Visually Impaired Gamers’ Preferences for Spatial Awareness Tools Within Video Games}. In \bibinfo{booktitle}{\emph{Proceedings of the 24th International ACM SIGACCESS Conference on Computers and Accessibility}}. \bibinfo{pages}{1--16}.
\newblock


\bibitem[Nair et~al\mbox{.}(2021b)]%
        {acoustic_game}
\bibfield{author}{\bibinfo{person}{Vishnu Nair}, \bibinfo{person}{Shao-en Ma}, \bibinfo{person}{Hannah Huddleston}, \bibinfo{person}{Karen Lin}, \bibinfo{person}{Mason Hayes}, \bibinfo{person}{Matthew Donnelly}, \bibinfo{person}{Ricardo~E Gonzalez}, \bibinfo{person}{Yicheng He}, {and} \bibinfo{person}{Brian~A. Smith}.} \bibinfo{year}{2021}\natexlab{b}.
\newblock \showarticletitle{Towards a Generalized Acoustic Minimap for Visually Impaired Gamers}. In \bibinfo{booktitle}{\emph{The Adjunct Publication of the 34th Annual ACM Symposium on User Interface Software and Technology}}. \bibinfo{publisher}{Association for Computing Machinery}, \bibinfo{address}{New York, NY, USA}, \bibinfo{pages}{89–91}.
\newblock
\showISBNx{9781450386555}
\urldef\tempurl%
\url{https://doi.org/10.1145/3474349.3480177}
\showURL{%
\tempurl}


\bibitem[Nair et~al\mbox{.}(2021c)]%
        {acousticmap}
\bibfield{author}{\bibinfo{person}{Vishnu Nair}, \bibinfo{person}{Shao-en Ma}, \bibinfo{person}{Hannah Huddleston}, \bibinfo{person}{Karen Lin}, \bibinfo{person}{Mason Hayes}, \bibinfo{person}{Matthew Donnelly}, \bibinfo{person}{Ricardo~E Gonzalez}, \bibinfo{person}{Yicheng He}, {and} \bibinfo{person}{Brian~A. Smith}.} \bibinfo{year}{2021}\natexlab{c}.
\newblock \showarticletitle{Towards a Generalized Acoustic Minimap for Visually Impaired Gamers}. In \bibinfo{booktitle}{\emph{Adjunct Proceedings of the 34th Annual ACM Symposium on User Interface Software and Technology}} (Virtual Event, USA) \emph{(\bibinfo{series}{UIST '21 Adjunct})}. \bibinfo{publisher}{Association for Computing Machinery}, \bibinfo{address}{New York, NY, USA}, \bibinfo{pages}{89–91}.
\newblock
\showISBNx{9781450386555}
\urldef\tempurl%
\url{https://doi.org/10.1145/3474349.3480177}
\showDOI{\tempurl}


\bibitem[Odom et~al\mbox{.}(2012a)]%
        {userenactment1}
\bibfield{author}{\bibinfo{person}{William Odom}, \bibinfo{person}{John Zimmerman}, \bibinfo{person}{Scott Davidoff}, \bibinfo{person}{Jodi Forlizzi}, \bibinfo{person}{Anind~K. Dey}, {and} \bibinfo{person}{Min~Kyung Lee}.} \bibinfo{year}{2012}\natexlab{a}.
\newblock \showarticletitle{A Fieldwork of the Future with User Enactments}. In \bibinfo{booktitle}{\emph{Proceedings of the Designing Interactive Systems Conference}} (Newcastle Upon Tyne, United Kingdom) \emph{(\bibinfo{series}{DIS '12})}. \bibinfo{publisher}{Association for Computing Machinery}, \bibinfo{address}{New York, NY, USA}, \bibinfo{pages}{338–347}.
\newblock
\showISBNx{9781450312103}
\urldef\tempurl%
\url{https://doi.org/10.1145/2317956.2318008}
\showDOI{\tempurl}


\bibitem[Odom et~al\mbox{.}(2012b)]%
        {userenactment3}
\bibfield{author}{\bibinfo{person}{William Odom}, \bibinfo{person}{John Zimmerman}, \bibinfo{person}{Scott Davidoff}, \bibinfo{person}{Jodi Forlizzi}, \bibinfo{person}{Anind~K. Dey}, {and} \bibinfo{person}{Min~Kyung Lee}.} \bibinfo{year}{2012}\natexlab{b}.
\newblock \showarticletitle{A Fieldwork of the Future with User Enactments}. In \bibinfo{booktitle}{\emph{Proceedings of the Designing Interactive Systems Conference}} (Newcastle Upon Tyne, United Kingdom) \emph{(\bibinfo{series}{DIS '12})}. \bibinfo{publisher}{Association for Computing Machinery}, \bibinfo{address}{New York, NY, USA}, \bibinfo{pages}{338–347}.
\newblock
\showISBNx{9781450312103}
\urldef\tempurl%
\url{https://doi.org/10.1145/2317956.2318008}
\showDOI{\tempurl}


\bibitem[Odom et~al\mbox{.}(2014)]%
        {userenactment2}
\bibfield{author}{\bibinfo{person}{William Odom}, \bibinfo{person}{John Zimmerman}, \bibinfo{person}{Jodi Forlizzi}, \bibinfo{person}{Hajin Choi}, \bibinfo{person}{Stephanie Meier}, {and} \bibinfo{person}{Angela Park}.} \bibinfo{year}{2014}\natexlab{}.
\newblock \showarticletitle{Unpacking the Thinking and Making behind a User Enactments Project}. In \bibinfo{booktitle}{\emph{Proceedings of the 2014 Conference on Designing Interactive Systems}} (Vancouver, BC, Canada) \emph{(\bibinfo{series}{DIS '14})}. \bibinfo{publisher}{Association for Computing Machinery}, \bibinfo{address}{New York, NY, USA}, \bibinfo{pages}{513–522}.
\newblock
\showISBNx{9781450329026}
\urldef\tempurl%
\url{https://doi.org/10.1145/2598510.2602960}
\showDOI{\tempurl}


\bibitem[Pavel et~al\mbox{.}(2020)]%
        {pavel2020rescribe}
\bibfield{author}{\bibinfo{person}{Amy Pavel}, \bibinfo{person}{Gabriel Reyes}, {and} \bibinfo{person}{Jeffrey~P. Bigham}.} \bibinfo{year}{2020}\natexlab{}.
\newblock \showarticletitle{Rescribe: Authoring and Automatically Editing Audio Descriptions}. In \bibinfo{booktitle}{\emph{Proceedings of the 33rd Annual ACM Symposium on User Interface Software and Technology}}. \bibinfo{publisher}{Association for Computing Machinery}, \bibinfo{address}{New York, NY, USA}, \bibinfo{pages}{747–759}.
\newblock
\showISBNx{9781450375146}
\urldef\tempurl%
\url{https://doi.org/10.1145/3379337.3415864}
\showURL{%
\tempurl}


\bibitem[Ramsdell(1978)]%
        {ramsdell1978psychology}
\bibfield{author}{\bibinfo{person}{Donald~A Ramsdell}.} \bibinfo{year}{1978}\natexlab{}.
\newblock \showarticletitle{The psychology of the hard-of-hearing and the deafened adult}.
\newblock \bibinfo{journal}{\emph{Hearing and deafness}}  \bibinfo{volume}{4} (\bibinfo{year}{1978}), \bibinfo{pages}{499--510}.
\newblock


\bibitem[Roo and Hachet(2017)]%
        {onereality}
\bibfield{author}{\bibinfo{person}{Joan~Sol Roo} {and} \bibinfo{person}{Martin Hachet}.} \bibinfo{year}{2017}\natexlab{}.
\newblock \showarticletitle{One Reality: Augmenting How the Physical World is Experienced by Combining Multiple Mixed Reality Modalities}. In \bibinfo{booktitle}{\emph{Proceedings of the 30th Annual ACM Symposium on User Interface Software and Technology}} (Qu\'{e}bec City, QC, Canada) \emph{(\bibinfo{series}{UIST '17})}. \bibinfo{publisher}{Association for Computing Machinery}, \bibinfo{address}{New York, NY, USA}, \bibinfo{pages}{787–795}.
\newblock
\showISBNx{9781450349819}
\urldef\tempurl%
\url{https://doi.org/10.1145/3126594.3126638}
\showDOI{\tempurl}


\bibitem[Rychtarikova(2015)]%
        {rychtarikova2015blind}
\bibfield{author}{\bibinfo{person}{Monika Rychtarikova}.} \bibinfo{year}{2015}\natexlab{}.
\newblock \showarticletitle{How do blind people perceive sound and soundscape}.
\newblock \bibinfo{journal}{\emph{Akustika}} \bibinfo{volume}{23}, \bibinfo{number}{1} (\bibinfo{year}{2015}), \bibinfo{pages}{6--9}.
\newblock


\bibitem[Sawhney and Schmandt(2000)]%
        {sawhney2000nomadic}
\bibfield{author}{\bibinfo{person}{Nitin Sawhney} {and} \bibinfo{person}{Chris Schmandt}.} \bibinfo{year}{2000}\natexlab{}.
\newblock \showarticletitle{Nomadic radio: speech and audio interaction for contextual messaging in nomadic environments}.
\newblock \bibinfo{journal}{\emph{ACM transactions on Computer-Human interaction (TOCHI)}} \bibinfo{volume}{7}, \bibinfo{number}{3} (\bibinfo{year}{2000}), \bibinfo{pages}{353--383}.
\newblock


\bibitem[Simeone et~al\mbox{.}(2015)]%
        {subsreality}
\bibfield{author}{\bibinfo{person}{Adalberto~L. Simeone}, \bibinfo{person}{Eduardo Velloso}, {and} \bibinfo{person}{Hans Gellersen}.} \bibinfo{year}{2015}\natexlab{}.
\newblock \showarticletitle{Substitutional Reality: Using the Physical Environment to Design Virtual Reality Experiences}. In \bibinfo{booktitle}{\emph{Proceedings of the 33rd Annual ACM Conference on Human Factors in Computing Systems}} (Seoul, Republic of Korea) \emph{(\bibinfo{series}{CHI '15})}. \bibinfo{publisher}{Association for Computing Machinery}, \bibinfo{address}{New York, NY, USA}, \bibinfo{pages}{3307–3316}.
\newblock
\showISBNx{9781450331456}
\urldef\tempurl%
\url{https://doi.org/10.1145/2702123.2702389}
\showDOI{\tempurl}


\bibitem[Siu et~al\mbox{.}(2020)]%
        {virtualcane}
\bibfield{author}{\bibinfo{person}{Alexa~F. Siu}, \bibinfo{person}{Mike Sinclair}, \bibinfo{person}{Robert Kovacs}, \bibinfo{person}{Eyal Ofek}, \bibinfo{person}{Christian Holz}, {and} \bibinfo{person}{Edward Cutrell}.} \bibinfo{year}{2020}\natexlab{}.
\newblock \showarticletitle{Virtual Reality Without Vision: A Haptic and Auditory White Cane to Navigate Complex Virtual Worlds}. In \bibinfo{booktitle}{\emph{Proceedings of the 2020 CHI Conference on Human Factors in Computing Systems}}. \bibinfo{publisher}{Association for Computing Machinery}, \bibinfo{address}{New York, NY, USA}, \bibinfo{pages}{1–13}.
\newblock
\showISBNx{9781450367080}
\urldef\tempurl%
\url{https://doi.org/10.1145/3313831.3376353}
\showURL{%
\tempurl}


\bibitem[Solutions(2023)]%
        {advancedhearing}
\bibfield{author}{\bibinfo{person}{Advanced~Hearing Solutions}.} \bibinfo{year}{2023}\natexlab{}.
\newblock \bibinfo{title}{How Does Active Noise Cancelling Work?}
\newblock
\newblock
\urldef\tempurl%
\url{https://hearlife.org/how-does-active-noise-cancelling-work/}
\showURL{%
\tempurl}


\bibitem[Stangl et~al\mbox{.}(2021)]%
        {stangl21assets}
\bibfield{author}{\bibinfo{person}{Abigale Stangl}, \bibinfo{person}{Nitin Verma}, \bibinfo{person}{Kenneth~R. Fleischmann}, \bibinfo{person}{Meredith~Ringel Morris}, {and} \bibinfo{person}{Danna Gurari}.} \bibinfo{year}{2021}\natexlab{}.
\newblock \showarticletitle{Going Beyond One-Size-Fits-All Image Descriptions to Satisfy the Information Wants of People Who Are Blind or Have Low Vision}. In \bibinfo{booktitle}{\emph{Proceedings of the 23rd International ACM SIGACCESS Conference on Computers and Accessibility}} (Virtual Event, USA) \emph{(\bibinfo{series}{ASSETS '21})}. \bibinfo{publisher}{Association for Computing Machinery}, \bibinfo{address}{New York, NY, USA}, Article \bibinfo{articleno}{16}, \bibinfo{numpages}{15}~pages.
\newblock
\showISBNx{9781450383066}
\urldef\tempurl%
\url{https://doi.org/10.1145/3441852.3471233}
\showDOI{\tempurl}


\bibitem[Tian et~al\mbox{.}(2021)]%
        {tian2021cyclic}
\bibfield{author}{\bibinfo{person}{Yapeng Tian}, \bibinfo{person}{Di Hu}, {and} \bibinfo{person}{Chenliang Xu}.} \bibinfo{year}{2021}\natexlab{}.
\newblock \showarticletitle{Cyclic co-learning of sounding object visual grounding and sound separation}. In \bibinfo{booktitle}{\emph{Proceedings of the IEEE/CVF Conference on Computer Vision and Pattern Recognition}}. \bibinfo{pages}{2745--2754}.
\newblock


\bibitem[Times(2020)]%
        {standardANC}
\bibfield{author}{\bibinfo{person}{The New~York Times}.} \bibinfo{year}{2020}\natexlab{}.
\newblock \bibinfo{title}{What Your Noise-Cancelling Headphones Can and Can’t Do}.
\newblock
\newblock
\urldef\tempurl%
\url{https://www.nytimes.com/wirecutter/blog/what-noise-cancelling-headphones-do/}
\showURL{%
\tempurl}


\bibitem[van Rijswijk and Strijbos(2013)]%
        {gpssound}
\bibfield{author}{\bibinfo{person}{Rob van Rijswijk} {and} \bibinfo{person}{Jeroen Strijbos}.} \bibinfo{year}{2013}\natexlab{}.
\newblock \showarticletitle{{Sounds in Your Pocket: Composing Live Soundscapes with an App}}.
\newblock \bibinfo{journal}{\emph{Leonardo Music Journal}}  \bibinfo{volume}{23} (\bibinfo{date}{12} \bibinfo{year}{2013}), \bibinfo{pages}{27--29}.
\newblock
\showISSN{0961-1215}
\urldef\tempurl%
\url{https://doi.org/10.1162/LMJ_a_00149}
\showDOI{\tempurl}
\showeprint{https://direct.mit.edu/lmj/article-pdf/doi/10.1162/LMJ\_a\_00149/1674871/lmj\_a\_00149.pdf}


\bibitem[Veluri et~al\mbox{.}(2023a)]%
        {veluri2023real}
\bibfield{author}{\bibinfo{person}{Bandhav Veluri}, \bibinfo{person}{Justin Chan}, \bibinfo{person}{Malek Itani}, \bibinfo{person}{Tuochao Chen}, \bibinfo{person}{Takuya Yoshioka}, {and} \bibinfo{person}{Shyamnath Gollakota}.} \bibinfo{year}{2023}\natexlab{a}.
\newblock \showarticletitle{Real-time target sound extraction}. In \bibinfo{booktitle}{\emph{ICASSP 2023-2023 IEEE International Conference on Acoustics, Speech and Signal Processing (ICASSP)}}. IEEE, \bibinfo{pages}{1--5}.
\newblock


\bibitem[Veluri et~al\mbox{.}(2023b)]%
        {semantichearing}
\bibfield{author}{\bibinfo{person}{Bandhav Veluri}, \bibinfo{person}{Malek Itani}, \bibinfo{person}{Justin Chan}, \bibinfo{person}{Takuya Yoshioka}, {and} \bibinfo{person}{Shyamnath Gollakota}.} \bibinfo{year}{2023}\natexlab{b}.
\newblock \showarticletitle{Semantic Hearing: Programming Acoustic Scenes with Binaural Hearables}. In \bibinfo{booktitle}{\emph{Proceedings of the 36th Annual ACM Symposium on User Interface Software and Technology}} (<conf-loc>, <city>San Francisco</city>, <state>CA</state>, <country>USA</country>, </conf-loc>) \emph{(\bibinfo{series}{UIST '23})}. \bibinfo{publisher}{Association for Computing Machinery}, \bibinfo{address}{New York, NY, USA}, Article \bibinfo{articleno}{89}, \bibinfo{numpages}{15}~pages.
\newblock
\showISBNx{9798400701320}
\urldef\tempurl%
\url{https://doi.org/10.1145/3586183.3606779}
\showDOI{\tempurl}


\bibitem[(W3C)(2021)]%
        {xraccessibiltiy}
\bibfield{author}{\bibinfo{person}{World Wide Web~Consortium (W3C)}.} \bibinfo{year}{2021}\natexlab{}.
\newblock \bibinfo{title}{XR Accessibility User Requirements}.
\newblock
\newblock
\urldef\tempurl%
\url{https://www.w3.org/TR/xaur/}
\showURL{%
\tempurl}


\bibitem[Wang et~al\mbox{.}(2022)]%
        {realitylens}
\bibfield{author}{\bibinfo{person}{Chiu-Hsuan Wang}, \bibinfo{person}{Bing-Yu Chen}, {and} \bibinfo{person}{Liwei Chan}.} \bibinfo{year}{2022}\natexlab{}.
\newblock \showarticletitle{RealityLens: A User Interface for Blending Customized Physical World View into Virtual Reality}. In \bibinfo{booktitle}{\emph{Proceedings of the 35th Annual ACM Symposium on User Interface Software and Technology}} (Bend, OR, USA) \emph{(\bibinfo{series}{UIST '22})}. \bibinfo{publisher}{Association for Computing Machinery}, \bibinfo{address}{New York, NY, USA}, Article \bibinfo{articleno}{49}, \bibinfo{numpages}{11}~pages.
\newblock
\showISBNx{9781450393201}
\urldef\tempurl%
\url{https://doi.org/10.1145/3526113.3545686}
\showDOI{\tempurl}


\bibitem[Wang et~al\mbox{.}(2020)]%
        {sliceoflight}
\bibfield{author}{\bibinfo{person}{Chiu-Hsuan Wang}, \bibinfo{person}{Chia-En Tsai}, \bibinfo{person}{Seraphina Yong}, {and} \bibinfo{person}{Liwei Chan}.} \bibinfo{year}{2020}\natexlab{}.
\newblock \showarticletitle{Slice of Light: Transparent and Integrative Transition Among Realities in a Multi-HMD-User Environment}. In \bibinfo{booktitle}{\emph{Proceedings of the 33rd Annual ACM Symposium on User Interface Software and Technology}} (Virtual Event, USA) \emph{(\bibinfo{series}{UIST '20})}. \bibinfo{publisher}{Association for Computing Machinery}, \bibinfo{address}{New York, NY, USA}, \bibinfo{pages}{805–817}.
\newblock
\showISBNx{9781450375146}
\urldef\tempurl%
\url{https://doi.org/10.1145/3379337.3415868}
\showDOI{\tempurl}


\bibitem[Zenner et~al\mbox{.}(2018)]%
        {immersivenoti}
\bibfield{author}{\bibinfo{person}{Andr\'{e} Zenner}, \bibinfo{person}{Marco Speicher}, \bibinfo{person}{S\"{o}ren Klingner}, \bibinfo{person}{Donald Degraen}, \bibinfo{person}{Florian Daiber}, {and} \bibinfo{person}{Antonio Kr\"{u}ger}.} \bibinfo{year}{2018}\natexlab{}.
\newblock \showarticletitle{Immersive Notification Framework: Adaptive \& Plausible Notifications in Virtual Reality}. In \bibinfo{booktitle}{\emph{Extended Abstracts of the 2018 CHI Conference on Human Factors in Computing Systems}} (Montreal QC, Canada) \emph{(\bibinfo{series}{CHI EA '18})}. \bibinfo{publisher}{Association for Computing Machinery}, \bibinfo{address}{New York, NY, USA}, \bibinfo{pages}{1–6}.
\newblock
\showISBNx{9781450356213}
\urldef\tempurl%
\url{https://doi.org/10.1145/3170427.3188505}
\showDOI{\tempurl}


\bibitem[Zhao et~al\mbox{.}(2018)]%
        {canetroller}
\bibfield{author}{\bibinfo{person}{Yuhang Zhao}, \bibinfo{person}{Cynthia~L. Bennett}, \bibinfo{person}{Hrvoje Benko}, \bibinfo{person}{Edward Cutrell}, \bibinfo{person}{Christian Holz}, \bibinfo{person}{Meredith~Ringel Morris}, {and} \bibinfo{person}{Mike Sinclair}.} \bibinfo{year}{2018}\natexlab{}.
\newblock \showarticletitle{Enabling People with Visual Impairments to Navigate Virtual Reality with a Haptic and Auditory Cane Simulation}. In \bibinfo{booktitle}{\emph{Proceedings of the 2018 CHI Conference on Human Factors in Computing Systems}}. \bibinfo{publisher}{Association for Computing Machinery}, \bibinfo{address}{New York, NY, USA}, \bibinfo{pages}{1–14}.
\newblock
\showISBNx{9781450356206}
\urldef\tempurl%
\url{https://doi.org/10.1145/3173574.3173690}
\showURL{%
\tempurl}


\bibitem[Zhao et~al\mbox{.}(2019)]%
        {seeingvr}
\bibfield{author}{\bibinfo{person}{Yuhang Zhao}, \bibinfo{person}{Edward Cutrell}, \bibinfo{person}{Christian Holz}, \bibinfo{person}{Meredith~Ringel Morris}, \bibinfo{person}{Eyal Ofek}, {and} \bibinfo{person}{Andrew~D. Wilson}.} \bibinfo{year}{2019}\natexlab{}.
\newblock \showarticletitle{SeeingVR: A Set of Tools to Make Virtual Reality More Accessible to People with Low Vision}. In \bibinfo{booktitle}{\emph{Proceedings of the 2019 CHI Conference on Human Factors in Computing Systems}}. \bibinfo{publisher}{Association for Computing Machinery}, \bibinfo{address}{New York, NY, USA}, \bibinfo{pages}{1–14}.
\newblock
\showISBNx{9781450359702}
\urldef\tempurl%
\url{https://doi.org/10.1145/3290605.3300341}
\showURL{%
\tempurl}


\bibitem[Zimmermann and Lorenz(2008)]%
        {zimmermann2008listen}
\bibfield{author}{\bibinfo{person}{Andreas Zimmermann} {and} \bibinfo{person}{Andreas Lorenz}.} \bibinfo{year}{2008}\natexlab{}.
\newblock \showarticletitle{LISTEN: a user-adaptive audio-augmented museum guide}.
\newblock \bibinfo{journal}{\emph{User Modeling and User-Adapted Interaction}} \bibinfo{volume}{18}, \bibinfo{number}{5} (\bibinfo{year}{2008}), \bibinfo{pages}{389--416}.
\newblock


\end{thebibliography}

\end{document}